\RequirePackage{fix-cm}
\documentclass[11pt,notitlepage,a4paper]{article}
\usepackage[a4paper,left=2.84cm,right=2.84cm,top=2.96cm,bottom=2.96cm,footskip=1.5cm]{geometry}
\renewcommand\footnotemark{}
\makeatletter
\renewcommand*{\@makefnmark}{}
\makeatother
\usepackage[hang,flushmargin]{footmisc}
\usepackage[hidelinks]{hyperref}
\usepackage{tocloft}

\newlength{\mylen}	
\usepackage{amsmath,amssymb,amscd,amsfonts}
\usepackage{enumitem}
\newcommand\aeq[1]{\begin{align}#1\end{align}}  
\newcommand\sigmaent{\sigma_{S}}
\newcommand\Sdensity{{\mathcal{S}}}
\newcommand\Real{{\mathbf{Re}}}
\newcommand\onlinecite{\cite}
\newcommand\kB{k}
\newcommand\crit{{\mathit{crit}}}
\newcommand{\dif}{\mathrm{d}}
\newcommand\rhoent{\rho_{S}}
\newcommand\eq{\begin{equation}}
\newcommand\en{\end{equation}}
\newcommand\expval[1]{\langle #1 \rangle}
\newcommand\expvalequil[1]{\expval{#1}_{\mathit{eq}}}
\newcommand\Jent{j_{S}}
\newcommand\Phient{\Phi_{S}}
\newcommand\Eent{E_{S}}
\newcommand\Vent{V_{S}}
\newcommand\comm[1]{{[}#1{]}}
\newcommand\eqa{\begin{eqnarray}}
\newcommand\ena{\end{eqnarray}}
\newcommand\cUV{c_{\mathit UV}}
\newcommand\JR{j_{R}}
\newcommand\JL{j_{L}}
\newcommand\Hzero{H_{0}}
\newcommand\calO{{\mathcal{O}}}
\newcommand\Tr{\mathrm{Tr}}
\newcommand{\me}{\mathrm{e}}
\newcommand\partialby[1]{\frac{\partial\hfill}{\partial#1}}
\newcommand\rhoequil{\rho_{\mathit{eq}}}
\newcommand\Yent{Y_{S}}
\newcommand\Edensity{{\mathcal{E}}}
\newcommand\Fdensity{{\mathcal{F}}}
\newcommand\Edensitycritical{{\mathcal{E}}_{\mathit crit}}
\newcommand\Sdensitycritical{{\mathcal{S}}_{\mathit crit}}
\newcommand\TR{T_{R}}
\newcommand\TL{T_{L}}
\newcommand\vev[1]{\langle 0| \, #1 \, |0\rangle}

\begin{document}
%
%
\thispagestyle{empty}
\begin{center}
{\LARGE Entropy Flow in Near-Critical Quantum Circuits
\footnotetext{In memory of Leo Kadanoff.}
\footnotetext{This is the first part of a two-part work originally published on 
arXiv.org in 
2005 as \cite{EntFlowI,DFEntropyFlowII}.
The second part follows in this volume \cite{DFEntropyFlowIIJStatPhys}.
For the present republication, some
revisions have been made for clarity, following helpful suggestions of the referee.
}}\\[6ex]
{\large Daniel Friedan}\\[1.5ex]
New High Energy Theory Center, Rutgers University, Piscataway, NJ, USA and \\\
Natural Science Institute, The University of Iceland, Reykjav\'\i k, Iceland\\
dfriedan@gmail.com\\[1.5ex]
February 21, 2017
\end{center}

\vspace*{2ex}

{\centering
\large\bfseries Abstract
\vskip1.5ex
}

Near-critical quantum circuits 
close to equilibrium
are ideal physical systems 
for asymptotically large-scale quantum computers, because 
their low energy collective excitations evolve reversibly, 
effectively isolated from microscopic environmental fluctuations
by the renormalization group.
Entropy flows in near-critical quantum 
circuits near equilibrium
as a locally conserved quantum 
current, obeying circuit laws analogous to the electric 
circuit laws. 
These ``Kirchhoff laws'' for entropy flow are the fundamental design 
constraints for asymptotically large-scale quantum computers.
A quantum 
circuit made from a near-critical system (of conventional 
type) is described by a relativistic 1+1 dimensional 
relativistic quantum field theory on the circuit.  
The quantum entropy current near equilibrium is just the 
energy current divided by the temperature.  
The 
universal properties of the energy-momentum tensor constrain 
the entropy flow characteristics of the circuit components: 
the entropic conductivity of the quantum wires and the 
entropic admittance of the quantum circuit junctions.  For 
example, near-critical quantum wires are always 
resistanceless inductors for entropy.  A universal formula 
is derived for the entropic conductivity: $\sigmaent(\omega) 
= iv^{2} \Sdensity /\omega T $, where $\omega$ is the 
frequency, $T$ the temperature, $\Sdensity$ the equilibrium 
entropy density and $v$ the velocity of ``light''.  The 
thermal conductivity is $\Real(T\sigmaent(\omega))=\pi v^{2} 
\Sdensity \, \delta(\omega)$.  The thermal Drude weight is, 
universally, $v^{2}\Sdensity$.  This gives a way to measure 
the entropy density directly.


\setcounter{tocdepth}{2}
\tableofcontents
\newpage
%
%
\section{Introduction}
Asymptotically large-scale quantum computers will have to 
operate reversibly, close to equilibrium, effectively 
isolated from the 
environment\cite{Landauer,Bennett,Benioff1,Benioff2}.  
Reversibility is a fundamental condition on any asymptotically 
large-scale computer, classical or quantum,
because irreversible computation necessarily generates heat.
A quantum computer has to maintain coherence as it evolves in 
time, so it must be isolated from environmental fluctuations.
In particular, it cannot afford to interact with the environment to 
discharge waste heat.  On the other hand, there must be some 
method of external control, for programming and for input 
and output.  External control requires contact with the 
environment.  A fundamental problem is to reconcile external 
control with isolation and reversibility.

Near-critical quantum systems close to equilibrium are ideal for the
purpose.  The low-energy collective excitations in a near-critical
quantum system are governed by a locally conserved energy density
operator.  This means that they form an isolated quantum system which
evolves reversibly in time.  The renormalizatgion group guarantees
that the low-energy excitations are effectively decoupled from
microscopic environmental influences.  External microscopic 
fluctuations affect the low-energy physics only
through a small number of relevant and marginal local couplings.
Control is feasible, in principle, because only the relevant and
marginal couplings need be tuned.  It might even be arguable that
near-critical quantum systems are the \emph{only} physical systems
that can operate reversibly and controllably at large scale, the only
physical systems in which asymptotically large-scale quantum
computation is practical.

This approach to the design of asymptotically large-scale 
quantum computers starts by singling out, on principle, the 
general class of useable physical systems.  Left for later 
is the question of how, precisely, quantum bits are to be 
represented and manipulated in near-critical quantum 
systems, on the perhaps facile assumption that a Hilbert 
space is, after all, just a Hilbert space.  When the low 
energy excitations are fermions, there is, of course, the 
obvious remark that the occupation number basis offers a 
quantum bit represention of the Hilbert space.

A quantum critical system is an extended system with a 
critical point at zero temperature.  When the couplings of 
the system approach their critical values, collective 
excitations develop whose energies and momenta are very 
small on the characteristic microscopic scales.  A 
``conventional'' quantum critical system is one in which, as 
the system approaches criticality, the energies and momenta 
of the low-energy excitations scale in the same way (see
\cite{Cardybook} and \cite{Sachdevbook}, 
for example).  The 
energy-momentum dispersion relation then takes the 
relativistic form $E(p)^{2} = E(0)^{2}+v^{2}p^{2}$, where 
$E(0)$ is the energy gap, which might be zero.  The entire low-energy 
physics becomes relativistic near the critical point, the 
coefficient $v$ being the speed of ``light''.  The low 
energy physics is described by a relativistic quantum field 
theory, whose scale of energy and momentum is very much 
smaller than the microscopic scale.  At low enough 
temperatures, such that $\kB T$ is small on the microscopic 
energy scale, the physics of the near-critical system is 
entirely due to the low-energy excitations, and is described 
by the relativistic quantum field theory at temperature $T$.  
Near-critical quantum circuits are one-dimensional 
near-critical systems, described by relativistic quantum 
field theories in 1+1 dimensions: one space dimension and 
one time dimension.  The ``conventional'' quantum critical 
systems are not the only kind of quantum critical system 
(see \cite{Sachdevbook} for example).  It might be 
that some of the present considerations apply as well to 
``non-conventional'' quantum critical systems, but only 
those described by relativistic quantum field theories will 
be considered here.
The uniform speed of ``light'' could be a distinguishing advantage 
of the relativistic quantum critical systems.

All near-critical quantum systems fall into a relatively 
small set of universality classes.  These are the asymptotic 
limits of renormalization.  The low-energy physics depends 
only on the universality class.  The physically possible 
universality classes correspond to mathematically possible 
relativistic quantum field theories, and it might be 
conjectured that every mathematically possible relativistic 
quantum field theory can be realized as a physical system.  
Each relativistic quantum field theory is parametrized by a 
finite number of renormalized coupling constants, and by the 
speed of ``light'', $v$, which is the maximum speed of the 
low-energy excitations.  All the low-energy properties of 
the near-critical quantum system are determined by the 
quantum field theory as universal functions of the 
renormalized coupling constants and $v$.  Each universality 
class can be implemented in a variety of microscopically 
different physical systems.  Nothing depends on the details 
of the physical implementation, except the value of $v$ and 
the map that takes the microscopic coupling constants in the 
physical system to the renormalized coupling constants of 
the quantum field theory.
  
The relativistic quantum field theories are the universal 
``machine languages'' for asymptotically large-scale quantum 
computers.  Large-scale quantum computers can be designed 
theoretically, in the language of quantum field theory.  If 
a particular universality class yields a theoretically 
useful design, the design can be implemented in any physical 
near-critical quantum system belonging to the universality 
class.  Algorithms for asymptotically large-scale quantum 
computation are to be designed within the constrained 
vocabulary of relativistic quantum field theory, rather than 
the much more general vocabulary of quantum mechanics.  The 
quantum field theory hamiltonians are very special among all 
possible quantum hamiltonians.  Within this highly 
constrained set of ``machine languages,'' it should be 
possible to make much more precise estimates of 
computational effectiveness than can be made for algorithms 
performed in general quantum mechanical systems.  General 
estimates should be derivable from general properties of 
quantum field theory, and more specific estimates from the 
particular properties of each individual universality class.

Large-scale quantum computers are likely to be built as 
circuits, for much the same reasons as classical computers.  
The basic reason is that one-dimensional circuits can be 
packed in three-dimensional space.  A more immediate reason 
is the expectation, or hope, that existing technologies for 
production of large-scale classical circuits can be adapted 
to production of large-scale quantum circuits.  A 
near-critical quantum circuit is described by a 1+1 
dimensional quantum field theory on the one-dimensional 
space of the circuit, a network of wires connected at 
junctions.  The 1+1 dimensional relativistic quantum field 
theories are the universal ``machine languages'' for 
near-critical quantum circuit computers.

The 1+1 dimensional quantum field theory determines the 
possible characteristics of the elementary circuit 
components, the quantum wires and junctions.  The challenge 
is to design circuits using these components that will 
perform useful large-scale quantum computations.  Only a 
preliminary step is taken here.  The laws governing entropy 
flow in near-critical quantum circuits
close to equilibrium 
are derived, and some
basic, elementary calculations are done.  Entropy is the 
currency of reversible computing.  The laws governing the 
flow of entropy are the basic constraints on the movement of 
information within a reversible computer, 
so the laws of entropy flow are the 
fundamental design constraints on
reversible computers.

For control of a near-critical quantum circuit to be 
practical, the wires will have to be exactly critical in the 
bulk, with no relevant bulk couplings at all.  The relevant 
couplings are the renormalizable couplings of positive 
scaling dimension.  Their effects become large at low 
energies and momenta.  Any relevant bulk coupling would have 
to be kept finely tuned everywhere along the length of the 
wires.  This would be a daunting task, probably hopeless.  
Quantum wires with no relevant bulk couplings will 
automatically be critical in the bulk.  No local 
perturbations will be able to make the bulk wire 
non-critical.  There might still be marginal bulk couplings 
to control, but that will not require fine tuning.  Such 
systems are gapless, and scale invariant in the bulk, and no 
local perturbation can produce a gap or disturb the scale 
invariance.  

The simplest example of a universality class without relevant 
operators is the 
Berezinskii-Kosterlitz-Thouless line of critical points in 
the $U(1)$-invariant 1+1 dimensional gaussian model.
There is a marginal scalar coupling that moves the system along the 
line of fixed points.
There are no relevant couplings that preserve the $U(1)$ 
symmetry.
But the model does have relevant charged operators.
So bulk scale invariance is not completely automatic.
Preserving the bulk scale invariance
requires screening against microscopic fluctuations
that violate the $U(1)$ symmetry.

The only known relativistic universality class that has no relevant 
operators whatsoever is 
the $c=24$ monster conformal field 
theory \cite{MONSTER1983,MONSTER1984,MONSTER1988}. 
The Monster 
conformal field theory
is the tensor product of two chiral 1+1 dimensional 
conformal field theories, each of which has the largest 
finite simple group, the monster group, as internal symmetry 
group.  This quantum field theory has no relevant bulk 
couplings and no scalar marginal couplings.
It does have a large number of spin-2 marginal couplings.
Assuming that the spin-2 marginal couplings can be controlled,
enforcing 1+1 dimensional relativistic symmetry,
the Monster conformal field theory is a fixed 
point of the renormalization group whose attracting basin is 
an open set in the space of physical systems.  To realize 
the monster field theory physically, all that is needed is a 
physical system whose couplings lie somewhere with the 
attracting basin of the monster fixed point.  No fine tuning
is needed at all.  On the other hand, the monster 
field theory is constructed mathematically as a chiral 
orbifold.  It is not clear how such a construction might be 
realized physically.  If the monster field theory could be 
realized in physical quantum wires, it would have distinct 
advantages over the gaussian model: no need to screen the 
wires to maintain a bulk $U(1)$ symmetry, 
nor any need to 
control a marginal scalar bulk coupling in the wires.  The low 
energy properties of the bulk wires would be completely 
insensitive to local perturbations.

The existence of such more or less
stable universality classes in 1+1 dimensions is another 
reason to prefer quantum circuits over quantum systems of 
higher dimension.
Ideal would be a universality class described by a 1+1 dimensional 
conformal field theory with no relevant or marginal 
operators at all, of any spin.  But no such theory has been found as 
yet.

All the relevant couplings of the near-critical quantum 
circuit will now be in the circuit junctions.  The junction 
couplings will represent the program and the input and 
output.  Computation will be performed by the time evolution 
of the quantum state of the low energy collective 
excitations traveling along the bulk-critical wires, 
scattering in the circuit junctions.  All excitations will 
travel at the speed of ``light'', $v$,
because of the bulk scale invariance.

Controlling the entropy of the junctions will be a crucial 
function in a near-critical quantum circuit computer.  
Entropy will be moved in and out of the junctions during the 
evolution of the circuit.  When the entropy of a junction is 
at its minimum, the junction is in a definite state, capable 
of supplying information, as program, or input, or output.  
When entropy flows into the junction, the state of the 
junction becomes uncertain, sensitive to the quantum 
excitations passing through.  When entropy flows out again, 
the junction is left in a new definite state, which is a 
function of the original state and of the excitations that 
passed through the junction.  The entropy flow 
characteristics of the circuit junctions will constrain the 
possible methods for input and output of information, and 
for control.

The junction entropy, $s$, was first described at critical 
points in the space of junction couplings, in the form 
$g=\exp(s_{\crit})$, the ``noninteger ground state 
degeneracy'' \cite{AL1}.  The junction is connected to long 
wires.  The junction entropy is what remains when the bulk 
entropy of the wires is subtracted from the total entropy of 
junction plus wires (with a subtlety: boundary conditions 
are needed for the far ends of the wires, whose entropies 
must also be subtracted).  When the junction is critical, 
along with the bulk wires, the whole system is scale 
invariant, so the junction entropy is independent of 
temperature, and can be attributed to the ground state.  It 
was conjectured in \cite{AL1}, and supported by 
considerable evidence, that the value of $g$ should always 
be larger at the ultraviolet critical point than at the 
infrared critical point, that it should decrease under the 
renormalization group flow, from fixed point to fixed point.  
This is the \emph{$g$-theorem}.

The $g$-theorem was proved by establishing a gradient 
formula, $\partial s /\partial \lambda^{a} = - 
g_{ab}(\lambda) \beta^{b}$, which expresses the variation of 
the junction entropy with respect to the junction coupling 
constants in terms of the junction beta-function, 
$\beta^{a}(\lambda)$, and a certain positive metric, 
$g_{ab}(\lambda)$, on the space of junction 
couplings \cite{FK}. The gradient formula holds for all 
junctions, critical or not.  The only scale in the junction 
is set by the temperature, $T$, so the renormalization group 
equation for the junction entropy is $T \dif s/\dif T = 
-\beta^{a} \partial s /\partial \lambda^{a}= \beta^{a}g_{ab} 
\beta^{b}$.  The junction entropy thus decreases with 
decreasing temperature, which is to say that it decreases 
under the renormalization group.  The junction contains 
minimum entropy at $T=0$, at its IR fixed point and maximum 
entropy at $T=\infty$, at its critical point, its UV fixed 
point.  The decrease of the junction entropy with 
temperature is not an obvious consequence of 
thermodynamics.  The junction is a bounded sub-system, but 
as part of a near-critical system it cannot be treated as 
finite.

It may well be useful that a junction is close to its 
critical point exactly when its entropy is close to its 
minimum.  Minimal entropy means that the junction is as 
sensitive as possible to the excitations passing through it.  
Close to critical means that the characteristic response 
time of the junction is long.  The combination seems ideal: 
the junction is then able to process over long times the 
effects of the excitations passing through it.

The gradient formula can be regarded as explaining how to 
control the junction entropy: how the junction entropy 
changes in response to changes in the junction couplings.  
By relating the change in entropy to the junction 
beta-function, the gradient formula suggests that control 
might be simplest to achieve in supersymmetric universality 
classes, since supersymmetry typically simplifies the 
renormalization group flow.  Supersymmetry is found at a 
special value of the marginal coupling constant in the 
$U(1)$-invariant gaussian 
model \cite{FQS1983,FQS1984,FS1986}.  Supersymmetry is also 
found in the $c=24$ monster conformal field 
theory \cite{dixon:1988}.

The junction entropy, $s(T)$, is not a useful quantity.  
Only \emph{changes}, $\Delta s(T)$, are physically 
significant, describing the movement of information in and 
out of the junction.  Of particular interest is the maximum 
change, $s(\infty)-s(0)$, which could be called the 
information capacity of the junction.  To find the junction 
entropy itself requires global measurement, but changes in 
the junction entropy can be determined locally, near the 
junction, by studying how entropy flows into and out from 
the junction.

\section{Mathematical setting}
The mathematical setting is as follows.
\begin{enumerate}
\item
There is a physical circuit described by
a 1+1 dimensional relativistic unitary quantum
field theory on  
a 1-dimensional space $\mathcal{C}$  with the geometry of a graph:
\begin{itemize}[label=\textbullet]
\item
a collection of line intervals (wires) and nodes (junctions),
\item each
boundary of a wire identified with one of the nodes.
\item
with space-time metric
$
g_{\mu\nu}dx^{\mu}dx^{\nu} = 
-v^{2}(dt)^{2}+(dx)^{2}
$,
\item
$v$ being the ``speed of light''.
\end{itemize}
\vskip1ex 

\item
There is a conserved, symmetric
energy-momentum tensor $T^{\mu\nu}(x)$
\begin{itemize}[label=\textbullet]
\item
which is an operator valued distribution on $\mathcal{C}$,
\item
which gives the hamiltonian
$
H_{0}= \int_{\mathcal{C}} dx \; T^{t}_{t}(x)
$.
\end{itemize}
\vskip1ex

\item
There are no additional symmetries in general (such as parity 
invariance).
\vskip1ex

\item
Bulk criticality is assumed
in sections~\ref{sect:bulkcriticalwire} and \ref{sect:chiralcurrents},
\begin{itemize}[label=\textbullet]
\item
implying bulk conformal invariance,
i.e., vanishing of the trace of the energy-momentum tensor
on the wires,
so the trace $T^{\mu}_{\mu}(x)$ consists entirely of delta-function contributions 
localized on the junctions.
\end{itemize}

\vskip1ex

\item
There is an equilibrium state at temperature $T = (k\beta)^{-1}$.
\begin{itemize}[label=\textbullet]
\item
Its partition function, density matrix, and expectation values are
\eq
\label{eq:equil1}
Z_{\mathit{eq}} = \Tr \left( e^{-\beta H_{0}}\right )
\,,\qquad
\rho_{\mathit{eq}} = 
\frac{e^{-\beta H_{0}}}{Z_{\mathit{eq}}}
\,,\qquad
\expvalequil{\mathcal{O}}
=\Tr \left ( \rho_{\mathit{eq}}\mathcal{O} \right )
\,.
\en
\end{itemize}

\vskip1ex

\item The system is assumed to operate close to the equilibrium state.

\item
Time-dependent perturbations of the hamiltonian
are studied
\eq
H(t) = H_{0} + \Delta H(t)
\en
of the form
\eq
\label{eq:pertH}
\Delta H(t) = \int_{\mathcal C} dx \;
T^{t}_{t}(x,t) \beta^{-1} \Delta \beta (x,t)
\en
\eq
\Delta\beta(x,t)\rightarrow 0  \quad\text{for } t\rightarrow 
\pm\infty 
\en
which amounts to perturbing by an inhomogeneous time-dependent 
temperature field,
\eq
\beta H(t) = 
\int_{\mathcal C} dx \;
T^{t}_{t}(x,t) \left [\beta + \Delta \beta (x,t)  \right ]
\,.
\en
The perturbation $\Delta H(t)$ is taken to be infinitesimal in order to 
derive linear response functions at equilibrium, but some basic results 
hold for arbitrarily large perturbations of the form (\ref{eq:pertH}).

\vskip1ex

\item
The departure of an observable from its equilibrium expectation value 
is written
\eq
\Delta {\mathcal O} = {\mathcal O} - \expvalequil{{\mathcal O}}
\,.
\en
\end{enumerate}

\section{Summary}

It is an elementary observation that entropy flows 
as a locally conserved quantum current
in near-critical quantum circuits that are close to equilibrium.
Every 1+1 dimensional quantum field theory 
has a conserved energy-momentum tensor, 
$T^{\mu}_{\nu}(x,t)$.  The energy density is 
$T^{t}_{t}(x,t)$, the energy current is $T^{x}_{t}(x,t)$.  
Close to equilibrium,
the entropy density operator, $\rhoent(x,t)$, is simply the 
change in the energy density from its equilibrium value, 
divided by the temperature
and
the entropy current operator $\Jent(x,t)$ is simply the energy current 
divided by the temperature,
\aeq{
\rhoent(x,t)&=\kB \beta 
T^{t}_{t}(x,t)
-\kB \beta  \expvalequil{T^{t}_{t}(x,t)}
\label{eq:rhoentsummary}
\\[1ex]
\Jent(x,t)&=\kB \beta 
T^{x}_{t}(x,t)
\:.
\label{eq:Jentsummary}
}
This is the local, quantum mechanical expression of the 
formula of Clausius, $\Delta S = \Delta Q/T$.  Local 
conservation of energy then implies local conservation of 
entropy,
\eq
\partial_{t} \rhoent(x,t) + \partial_{x} \Jent(x,t) = 0
\:.
\en
The idea that entropy moves locally within 
a compound system goes back at least to Gibbs \cite{Gibbs}.
What is perhaps slightly novel here is the construction of 
the entropy density as a \emph{quantum} operator.

The flow of entropy can now be treated in analogy with the 
flow of electric charge in electric circuits.  Entropic 
circuit laws can be written, in analogy with Kirchhoff's laws for
electric circuits.  The entropic potential, $\Phient(x,t)$,
is defined in analogy with the electric potential, as
a source that couples to the entropy density, perturbing the 
hamiltonian by
\eq
\Delta H(t) =
\int dx \; \rhoent(x,t) \Phient(x,t)
= \int dx \; T^{t}_{t}(x,t) k\beta \Phient(x,t)
\en
so the entropic potential can be interpreted as
a local variation of the temperature,
\eq
\Phient(x,t) = T^{-2} \Delta(T^{-1})(x,t)\,.
\en
In the linearized approximation, the entropic potential
is the local temperature drop
\eq
\Phient(x,t) 
\approx -\Delta T(x,t)\,.
\en
The entropic field, analogous to the electric field, is defined as 
the gradient of the entropic potential,
\eq
\Eent(x,t) = - \partial_{x} \Phient(x,t)
\,.
\en
It can be interpreted as the temperature gradient
\eq
\Eent(x,t) 
\approx
\partial_{x} T(x,t)
\,.
\en
The laws governing entropy flow are more stringent than the 
laws for electricity, because the energy current is part of 
the conserved, symmetric energy-momentum tensor, 
$T^{\mu\nu}=T^{\nu\mu}$.  The symmetry of the 
energy-momentum tensor relates the energy current to the 
momentum density: $T^{x}_{t}(x,t) = -v^{2} T^{t}_{x}(x,t)$.  
This significantly constrains the entropy characteristics of 
a quantum circuit.

The flow of entropy in a wire is characterized by its entropic
conductivity.  The entropic conductivity of a wire,
$\sigmaent(\omega)$, is the linear response coefficient which gives
the entropy current, $\Delta I_{S}(t) = \sigmaent(\omega) \Delta
\Eent(t)$, that flows in response to an infinitesimal uniform entropic
field alternating at frequency $\omega$.  

The flow of entropy through a junction is characterized by its
entropic admittance, which gives the
entropy current, $\Delta I_{S}(t)_{A} = \sum_{B=1}^{N}
Y_{S}(\omega)_{AB}\,\Delta \Vent(t)^{B}$, that flows out each of the
wires connected to the junction, $A=1\ldots N$, in response to small
alternating changes, $\Delta \Vent(t)^{B}$, in the entropic potentials
on the wires.

The entropic circuit laws determine the entropy flow in a complex
circuit from the entropy characteristics of its components.  A complex
circuit is a collection of quantum junctions connected by quantum wires.  
Each junction is either elementary, without substructure,
or is a complex sub-circuit,
whose sub-structure is expressed in the entropic admittance
$Y_{S}(\omega)_{AB}$
of the junction.

This paper defines the entropy density and current 
operators, notes that circuit laws for entropy follow 
by formal analogy to the electric circuit laws, derives
entropy continuity and conduction equations, and a
universal formula for the entropic conductivity.  All the 
derivations and calculations are completely elementary.  A 
subsequent paper treats the flow of 
entropy through near-critical quantum circuit junctions 
\cite{DFEntropyFlowII,DFEntropyFlowIIJStatPhys}.

In reversible processes close to equilibrium, bookkeeping
entropy is equivalent to bookkeeping energy.
The entropy density and current operators
defined here are just the energy density and current
operators re-scaled.
The present results on entropy transport can all be re-phrased as
results on energy transport by simply re-scaling the formulas.  
It is
quite standard to express thermal transport properties of quantum
systems in terms of the local energy density and current operators
\cite{luttinger:1964}.

However, it is entropy transport that is of
interest in reversible computing, since entropy transport
is the negative of information transport.
Moreover, it is the entropy current and density operators that
are appropriate for the formulation of circuit laws, because the
entropic potential, $\Phient(x,t)$ has a thermodynamic interpretation
as the local drop in temperature, $\Phient(x,t)=-\Delta T(x,t)$.  No
such thermodynamic quantity couples to the energy density.

Much of what is done here is related to other works.  Quantum wires
and junctions have been much studied (see, for example,
\onlinecite{QuantumCircuits8} and references therein).  The idea of
using quantum critical systems for quantum computers is far from new
(see, for example, \onlinecite{Sachdevarticle}).  But the goal and the
reasoning are perhaps different here.  The goal here is to estalish
the fundemental physical constraints on the design of asymptotically
large-scale quantum computers, based on the condition that such
computers must be large quantum systems that (1) operate reversibly near
equilibrium and (2) are controllable.  The basic reason for proposing that
asymptotically large-scale quantum computers should be built from
near-critical quantum circuits is the effective isolation provided by
renormalization.  Renormalizaton, by decoupling the low energy
excitations from the microscopic physics, solves the hard part of the
control problem andthus  makes reversible operation practical.  The program
that follows from this reasoning is to base the design strategy on the
laws of entropy flow.

Universal equal-time commutation relations are derived for
the energy density operator $T^{t}_{t}(x,t)$ and the energy
current operator $T^{x}_{t}(x,t)$.
The formula for the equal-time commutator
$\comm{T^{t}_{t}(x',t),T^{t}_{t}(x,t)}$
implies the continuity equation for entropy:
\eq
\begin{split}
\partial_{t} \expval{\rhoent(x,t)}
+\partial_{x}\expval{\Jent(x,t)}
&=
\kB \beta \Eent(x,t)  \expval{\Jent(x,t)}
\\
&\qquad - 
\kB \beta \partial_{x} 
\left [
\Phient(x,t)
\expval{  \Jent(x,t)}
\right ]
\:.
\label{eq:continuity}
\end{split}
\en
The formula for $\comm{T^{t}_{t}(x',t),T^{x}_{t}(x,t)}$
implies an equation for entropy conduction in bulk wires:
\eq
\begin{split}
\partial_{t}\expval{\Jent(x,t)} &=
- 
\kB^{2} \beta^{2} v^{2} 
\frac{\cUV}{6} \frac{\hbar v}{2\pi} \partial_{x}^{2}\Eent(x,t)
\\
&\qquad\qquad {} +
\kB^{2} \beta^{2} v^{2}
\expval{T_{t}^{t}(x,t)-T_{x}^{x}(x,t)}
\Eent(x,t)
\\
& \qquad \qquad
{}
+
\left [
1
+\kB \beta \Phient(x,t)
\right ]
\kB\beta v^{2}
\partial_{x}\expval{T_{x}^{x}(x,t)}
\:.
\end{split}
\label{eq:conduction}
\en
where $\cUV$ is the bulk conformal central charge in the 
short-distance limit (for discussion of 1+1 
dimensional conformal field theory, see, for example, 
\onlinecite{Cardybook} 
and~\onlinecite{DiFran}).  

The continuity and 
conduction equations are exact.  No linear response 
approximations are made.
They are better interpreted as
continuity and conduction equations for 
energy,
by substituting for $\rhoent$ and $\Jent$ 
according to (\ref{eq:rhoentsummary})
and (\ref{eq:Jentsummary}).
This is especially the case for the conduction equation (\ref{eq:conduction}),
which is not expressed entirely in terms of $\rhoent$ and $\Jent$,
but also involves other components of the energy-momentum tensor.

The conduction equation is useful in the limit of uniform 
flow, when the spatial derivatives can be neglected, and in 
the bulk-critical limit, where conformal invariance allows 
replacing $T_{x}^{x}(x,t)$ with $- T_{t}^{t}(x,t)$.  The 
conduction equation, in the linear response approximation 
and the uniform limit, implies a general formula for the 
entropic conductivity of near-critical quantum wire:
\eq
\sigmaent(\omega)= 
\frac{i v^{2} \Sdensity}{\omega T}
\label{eq:conductivity}
\en
where $\omega$ is the frequency, $T$ is the temperature, 
$\Sdensity$ is the equilibrium entropy density, and $v$ the 
velocity of ``light''.  Thus near-critical quantum wires, as 
circuit elements, are resistanceless inductors for entropy.
A direct
derivation of (\ref{eq:conductivity})
using the Kubo formula
is given in Appendix~\ref{app:kubo}.
The entropic conductivity is just the complex thermal conductivity 
divided by temperature, so the thermal conductivity of 
near-critical quantum wire is
\eq
\kappa(\omega,T)  = \Real( T \sigmaent(\omega)) =
\pi v^{2} \Sdensity \,\delta(\omega)
\label{eq:thermalconductivity}
\:.
\en
The coefficient $v^{2}\Sdensity$ is the universal thermal 
Drude weight for a near-critical one-dimensional quantum 
system.

The universal formula for the entropic conductivity of 
quantum wire that is near-critical but not critical is not 
directly relevant to the design of quantum circuit 
computers, given the argument that the quantum wires in such 
circuits should be exactly critical in the bulk.  On the 
other hand, the universal formula (\ref{eq:thermalconductivity})
for the thermal conductivity could 
be useful in other contexts, as it gives a way to determine 
directly, by experiment, the entropy density of the low 
energy collective excitations in near-critical 
one-dimensional and quasi-one-dimensional quantum systems.

When the quantum wires are critical in the bulk,
the entropic conductivity
for non-uniform flow is calculated
from the continuity and conduction equations,
in the linear response approximation,
using bulk conformal invariance:
\eq
\sigmaent(q,\omega) 
=
\frac{c}{12}
\frac{2\pi\kB^{2}v}{\hbar}
\left [ 1
+ \left ( \frac{\hbar v\beta q}{2\pi} \right ) ^{2}
\right ]
\left (
\frac{i}{\omega+i\epsilon+vq}
+\frac{i}{\omega+i\epsilon-vq}
\right )
\label{eq:critcond}
\en
where $c$ is the bulk conformal central charge and $q$ is 
the wave-number.  
As is very well-known in 1+1 dimensional conformal field theory,
the energy current splits into independent right-moving and 
left-moving chiral energy currents.
The low-energy excitations all travel at 
the speed of ``light''.
The 
entropy current is just the energy current divided 
by the temperature,
so the entropy current 
splits into 
independent right-moving and 
left-moving chiral entropy currents,
$\JR(x,t)=\JR(x-vt)$ and $\JL(x,t)=\JL(x+vt)$.
This is exhibited
in equation~(\ref{eq:critcond}) by the poles in 
$\sigmaent(q,\omega) $ at $\omega=\pm v q$.  
The chirality of the entropy currents 
--- the uniform speed of the excitations and the absence of 
bulk interactions between the left-moving and right-moving 
entropy currents --- might be a useful feature of 
bulk-critical quantum wires.

Formulas for the thermal conductivity equivalent to 
(\ref{eq:thermalconductivity}) were previously 
derived for the special case of a 1d spin chain \cite{KlumperSakai}
and the special cases of free massive fermions and 
bosons and for the general bulk-critical 
case \cite{orignac:134426}, but it was not noted for any of 
these special cases that the thermal Drude weight takes the general
form $v^{2}\Sdensity$.

\section{Comments on the occasion of republication}

This paper and its sequal were originally published on 
arXiv.org in 2005 \cite{EntFlowI,DFEntropyFlowII}.
In the present version, the author has made a number of revisions for 
clarity following suggestions of the referee
of the present version,
but has not changed anything of substance,
with two minor exceptions:
(1) deletion of the argument on the temperature dependence of persistent equilibrium 
entropy currents that appeared in the original version of the first paper at the end 
of section VI (where the argument was described as ``dubious''),
and (2) addition of a comment in the Introduction above on the problem of 
marginal spin-2 operators in the $c=24$ monster conformal field 
theory.

The referee of the present version suggested adding some
discussion of how these two papers fit into the modern research on quantum
states out of equilibrium, pointing out that ``a lot of work has been
done since these papers first appeared on the arxiv'', and pointing 
for possible references to the various reviews in 
\cite{Calabreseetal}.
The author takes the position that the two papers from 2005 are
here being republished in honor of Leo Kadanoff
and that discussion of work after the original 2005 publication on 
arXiv.org would be ahistorical.
The author also takes the position that he would not be competent to 
discuss the work that has taken place since the original publication.
The reviews in \cite{Calabreseetal}, and especially \cite{Doyon},
give very many references, if perhaps not all.

In any case, research on out of equilibrium physics is not germaine.
The starting point of the present work is the fact that asymptotically
large scale computers, classical or quantum, must operate reversibly,
close to equilibrium \cite{Landauer,Bennett,Benioff1,Benioff2}.  The
fundamental physical constraints on such computers are the laws of
local entropy flow close to equilibrium, which are given by linear
response theory, in analogy with Kirchoff's laws for electric
circuits.  Some of the results of the present papers apply to energy
flow out of equilibrium, but these results are incidental.

The review  \cite{Doyon} gives some
references to relevant work before the original 2005 publication
of the present papers,
that the author was ignorant of at the time.
In particular, both \cite{MacLennan} and \cite{Zubarev} construct a classical entropy density 
and entropy current and write a local conservation law for entropy.
Still, the attribution of the original idea to Gibbs \cite{Gibbs} at the beginning of the 
Summary above seems correct.
Several of the pre-2005 papers cited in \cite{Doyon} discuss flow of 
energy and/or flow of information and entropy,
but not in circumstances related to the present work.
Several discuss a 1d system (a wire)  in a ``non-equilibrium steady state''
with different temperatures at its two 
boundaries.
Again, some of the present results on energy transport and 
locally varying temperature might be applicable to such an out of 
equilibrium situation,
but the central concern here is only with small local perturbations around 
the equilibrium state.
In particular, it should be noted
that the persistent equilibrium energy current $\expvalequil{\Jent(x,t)}$.
discussed at the end of section \ref{sect:continuity} below is not the same as the 
entropy current in the ``non-equilibrium steady state''.
The latter is steadily pumping entropy into the wire, thus 
``non-equilibrium''.
Finally, the paper \cite{KlumperSakai} calculated the thermal conductivity in 
a 1d spin chain.   It has now been added to the citation of 
\cite{orignac:134426} at the end of the Summary above.

\section{Entropy flow}

Every relativistic quantum field theory has a locally 
conserved, symmetric energy-momentum tensor, 
$T_{\nu}^{\mu}(x,t)$, which represents the response of the 
system to local deformations of the space-time geometry.  In 
1+1 dimensions, the space-time metric, $g_{\mu\nu}$, has 
components $g_{tt}=-v^{2}$, $g_{xx}=1$, $g_{xt}=g_{tx}=0$.  
Any local deformation, $\delta g_{\mu\nu}(x,t)$, can be 
represented as a local variation of the couplings in the 
hamiltonian, so is equivalent to a perturbation of the 
hamiltonian by a local quantum field, the energy-momentum 
tensor.  The expectation values in the quantum field 
theory change by
\eq
\delta \expval{\cdots} = \int \int \dif t \, \dif x
\,\,  \left (\frac{-1}{2}\right )
\delta g_{\mu\nu}(x,t) \,
\expval{\frac{i}{\hbar} T^{\mu\nu}(x,t) \cdots }
\en
where
$T^{\mu\nu'}g_{\nu'\nu} = T^{\mu}_{\nu}$.
The energy-momentum tensor is symmetric,
$T^{\mu\nu}=T^{\nu\mu}$,
because the space-time metric is symmetric.
Symmetry of $T^{\mu\nu}$
is simply the indentity $T^{x}_{t} = -v^{2} T^{t}_{x}$.

The individual components of the energy-momentum tensor are 
the energy density, $T_{t}^{t}(x,t)$, the energy current, 
$T_{t}^{x}(x,t)$, the momentum density, $T_{x}^{t}(x,t)$, and 
the momentum current, $T_{x}^{x}(x,t)$.
Energy and momentum are each locally conserved,
\eq
\partial_{\mu} 
T_{\nu}^{\mu}(x,t) = 0
\:.
\en
The hamiltonian is 
\eq
\label{eq:hamiltonian}
\Hzero = \int \dif x \,\, T_{t}^{t}(x,t)
\:.
\en
Local conservation of energy,
\eq
\partial_{t}T_{t}^{t}(x,t) + \partial_{x}T_{t}^{x}(x,t) = 0
\:,
\en
expresses the effective isolation of the near-critical 
quantum system.  The degrees of freedom of the quantum field 
theory are the low energy collective degrees of freedom of 
the near-critical quantum system.  They form a closed 
system, neither gaining nor losing energy, effectively 
decoupled from microscopic fluctuations.

The formula of Clausius,
\eq
\Delta S  = \frac{\Delta Q}{T}
\:,
\label{eq:reversible}
\en
expresses the change of entropy in a reversible process as 
the change in heat divided by the temperature.  The constant 
of proportionality, $1/T$, is also written $\kB \beta$, 
where $\kB$ is Boltzmann's constant, the fundamental unit of 
entropy.  In a near-critical system, the change in heat 
within a local region, ${\mathcal{R}}$, is
\eq
\Delta Q_{\mathcal{R}}= 
\int_{\mathcal{R}} \dif x \,\,\Delta \expval{T_{t}^{t}(x,t)}
\en
because all available forms of energy are
included in $T_{t}^{t}(x,t)$.
The entropy within region ${\mathcal{R}}$ changes by
\eq
\Delta S_{\mathcal{R}} = 
\kB  \beta \,
\int_{\mathcal{R}} \dif x \,\, \Delta \expval{T_{t}^{t}(x,t) }
\:.
\en
This local version of the Clausius formula can be derived 
formally by calculating the change of entropy when the 
hamiltonian of the local region is perturbed infinitesimally 
from $\Hzero$ to $H=\Hzero+\Delta H$, at constant 
temperature.  The equilibrium expectation values of observables, 
$\expvalequil{\calO}$, are perturbed to 
$\expval{\calO}=\expvalequil{\calO} +\Delta \expval{\calO}$.  
The partition function is
\eq
Z = \Tr(\me^{-\beta H})
\en
and the entropy is
\eq
S = \kB  \left (1 - \beta \partialby{\beta} \right ) \ln Z
= \kB \ln Z + \kB\beta\expval{ H}
\:.
\en
The infinitesimal change of entropy is
\eqa
\Delta S  &=& \kB \Delta \ln Z
+ \kB\beta \Delta\expval{\Hzero} + \kB\beta\expvalequil{ \Delta H}
\nonumber \\
&=& \kB \beta  \expvalequil{-\Delta H}
+ \kB\beta \Delta\expval{ \Hzero} + \kB\beta\expvalequil{ \Delta H}
\nonumber  \\
&=& \kB\beta \Delta\expval{ \Hzero}
\nonumber  \\
&=& \kB\beta \Delta\expval{
\int \dif x \,\, T_{t}^{t}(x,t)}
\:.
\ena
The Clausius formula for finite perturbations
follows by
integrating
over infinitesimal perturbations.

Write the local change of entropy as
\eq
\Delta S_{\mathcal{R}} = \Delta \int_{\mathcal{R}}\dif x \,\,
\expval{ \rhoent (x,t) }
\en
where
\eq
\label{eq:rhoent}
\rhoent (x,t)
= \kB  \beta T_{t}^{t}(x,t)
  - \expvalequil{\kB  \beta T_{t}^{t}(x,t)}
\:.
\en
The operator $\rhoent (x,t)$
can be interpreted as the variation of the entropy density
away from its
equilibrium value,
which is the natural baseline against 
which to measure the flow of entropy.
The time derivative of the entropy density operator is
\eq
\partial_{t} \rhoent(x,t)
=\kB  \beta \partial_{t} T_{t}^{t}(x,t)
= -\kB \beta \partial_{x} 
T_{t}^{x}(x,t)
\en
so the local entropy current is
\eq
\label{eq:Jent}
\Jent(x,t) =  \kB  \beta T_{t}^{x}(x,t) 
\:.
\en
The entropy current, $\Jent(x,t)$, is the entropy per unit time flowing
to the right through the point $x$.
The local entropy current and the local 
entropy density, $\rhoent(x,t)$, are \emph{quantum} 
observables.
Entropy flows as a locally conserved \emph{quantum} current:
\eq
\partial_{t} \rhoent(x,t) + \partial_{x} \Jent(x,t) = 0
\:.
\en

\section{Quantum circuits and entropic circuit laws}

A near-critical quantum circuit is described by a 1+1 
dimensional relativistic quantum field theory on the 
one-dimensional space of the circuit.  That space consists 
of a set of line segments, the wires, and a set of points, 
the junctions.  Each wire boundary is identified with one of 
the junctions.  A junction to which only a single wire is 
connected is simply an end of the wire.

The existence of the locally conserved quantum entropy 
current implies that the flow of entropy is governed by 
entropic circuit laws derived by formal analogy with the 
laws of electric circuits, which are expressed for example 
in Kirchoff's laws.  The circuit laws determine the 
performance of the whole circuit given the characteristics 
of its parts, the entropic conductivity of the wires and the 
entropic admittance of the junctions.

The characteristics of the wires and junctions are 
determined by their linear responses to small external 
perturbations by an entropic potential, $\Phient(x,t)$, 
which is the external source that couples to the entropic 
charge density, in analogy with the electric potential.
The perturbed hamiltonian is
\eq
H = \Hzero + \Delta H(t) = \Hzero + \int \dif x \,\,  \rhoent(x,t)
\Phient(x,t)
\:.
\en
$\Phient(x,t)$ has dimensions of temperature.  The entropic 
field
\eq
\label{eq:entropicfield}
\Eent(x,t) =  - \partial_{x}  \Phient(x,t)
\en
has dimensions of temperature/distance.

Compare this perturbation to a change of
temperature
from $T$ to $T+\Delta T$.
The equilibrium density matrix of the unperturbed system is
\eq
\rhoequil = \frac1{Z_{0}} \me^{-\beta \Hzero}
\:.
\en
Under an infinitesimal change of temperature,
the equilibrium density matrix changes by
\eq
\Delta \rhoequil = 
\rhoequil
( \kB \beta^{2}  \Delta T) \;
(\Hzero -\expvalequil{\Hzero})
\:.
\en
The same effect
can be obtained
at constant temperature 
by adding an infinitesimal perturbation
to the
hamiltonian,
\eq
\Delta H =   (\Hzero -\expvalequil{\Hzero}) (- \kB  \beta \Delta T)
= \int \dif x \,\, \rhoent(x,t)(- \Delta T)
\:.
\en
Therefore, imposing an infinitesimal static entropic 
potential, $\Delta \Phient(x,t)=-\Delta T(x)$, is equivalent 
to making an infinitesimal local variation of the 
temperature, $\Delta T(x)$, in the limit where both become
constant in space.  Integrating infinitesimal 
perturbations gives $\Delta 
\Phient(x,t)=-\Delta T$ for finite changes of temperature.  
Increasing the entropic potential means \emph{decreasing} 
the temperature.  The entropic potential is the local \emph{drop} 
in temperature.  The entropic field $\Eent(x,t) = - 
\partial_{x} \Phient(x,t)$ is the local temperature 
gradient, in the limit where both are uniform in 
space.

When the external perturbation is turned on, entropy flows 
along the entropic field from higher entropic potential to 
lower, from regions of \emph{lower} temperature to regions 
of \emph{higher} temperature.  The entropic potential acts 
like the temperature dial on the thermostat of a heating 
system.  When a negative entropic potential is introduced in 
a local region, the temperature dial there is turned 
\emph{up}.  The couplings in the hamiltonian are changed 
locally, so that the system behaves as if at a higher 
temperature, locally.  The system responds by evolving 
towards local equilibrium at the new temperature.  
Initially, there is too little entropy in the perturbed 
region for that region to be in equilibrium at the new, 
higher temperature, so entropy flows \emph{into} the 
perturbed region from regions of higher entropic potential 
elsewhere in the system.

In an operating circuit, the only external perturbations 
will be in the junctions that serve as external controls.  
The entropic potential everywhere else in the circuit is an 
auxiliary variable, determined by a subset of the circuit 
equations as a function of the entropic currents and 
charges.  The remaining circuit laws and the external 
entropic potentials in the control junctions then determine 
the flow of entropy within the circuit.

\section{Entropic conductivity and admittance}

The entropic conductivity of a quantum wire is defined by
analogy with the electrical conductivity.
When a wire is perturbed by a small
entropic field, $ \Delta \Eent(x,t) = \me^{iqx-i\omega t}
\Delta \Eent(0,0) $, an entropy current is induced to flow.  The 
entropy current is given, to first order in the 
perturbation, by a linear response formula
\eq
\label{eq:linearresponse}
\Delta \expval{\Jent(x,t)}
= \sigmaent(q,\omega) \Delta \Eent(x,t)
\:.
\en
The linear response coefficient $\sigmaent(q,\omega)$
is the entropic conductivity.
Appendix~\ref{app:kubo} gives the usual derivation of the linear 
response formula (\ref{eq:linearresponse}) for the perturbed current
and the usual derivation of the Kubo formula for the conductivity
in terms of two-point functions.
The entropic conductivity for uniform flow is
\eq
\sigmaent(\omega) =
\lim_{q\rightarrow 0} 
\sigmaent(q,\omega)
\:.
\en
The entropic conductivity is just the complex thermal conductivity 
divided by the temperature,
\eq
\sigmaent = T^{-1}\sigma_{\mathit{thermal}}
\,,
\en
since $\Delta \Eent$ is the temperature 
gradient and $\Delta \expval{\Jent}$ is the energy current divided by 
temperature.

The circuit laws for uniform entropy conduction through wires are 
analogous to those for electrical conduction.
Let $A$,$A'$ label the two ends of a wire.
Let $\Delta I_{S}(t)_{A}$ be the entropy current entering the wire at 
end $A$,
and $\Delta I_{S}(t)_{A'}$ the entropy current entering at end $A'$.
Let $\Delta \Vent(t)^{A}$ be
the entropic potential at end $A$,
and $\Delta \Vent(t)^{A'}$
the entropic potential at end $A'$.
Let $\Delta E_{S}(t)$ be the entropic field in the wire.
Let $l$ be the length of the wire.
The conduction equations for uniform
entropy flow in the wire are
simple consequences of (\ref{eq:entropicfield}) and (\ref{eq:linearresponse}),
\eqa
l\,\Delta E_{S}(t)&=& \Delta \Vent(t)^{A}-
\Delta \Vent(t)^{A'}\\
\Delta I_{S}(t)_{A}
&=& \sigmaent(\omega) \Delta E_{S}(t)
= - \Delta I_{S}(t)_{A'}
\:.
\ena

The entropic admittance of a junction
is defined by analogy with the electrical
admittance.
Label the
$N$ wire-ends attached to the junction
by indices $A,B=1\ldots N$.
Let $\Delta \Vent(t)^{B}$ be
an infinitesimal change of
entropic potential at the end of wire $B$,
where it is attached to the junction,
and let $\Delta I_{S}(t)_{A}$
be the resulting change in the
entropic current flowing out of the junction
through wire $A$.
For alternating potentials
\eq
\Delta \Vent(t)^{B} = \me^{-i\omega t} \; \Delta \Vent(0)^{B}
\en
the junction admittance equation is
\eq
\Delta I_{S}(t)_{A} = \sum_{B=1}^{N}
\Yent(\omega)_{AB}\,\Delta \Vent(t)^{B}
\label{eq:junction}
\en
where the matrix $\Yent(\omega)_{AB}$ is the 
entropic admittance of the junction.
The usual derivation of the linear response formula (\ref{eq:junction}) and 
of the Kubo formula for the admittance 
can be found in section 2 of the second paper
\cite{DFEntropyFlowIIJStatPhys}.

The entropic conductivity, $\sigmaent(\omega)$, of the wire 
and the entropic admittance matrices, $\Yent(\omega)_{AB}$, 
of the elementary junctions are to be calculated in the 1+1 
dimensional quantum field theory.  The conduction equations 
for the wires and the admittance equations for the junctions 
then determine the entropy flow properties of the circuit.

\section{The continuity equation for entropy}
\label{sect:continuity}

Observables in the unperturbed system evolve in time by
\eq
\partial_{t} \calO(t)
=\frac{i}\hbar \comm{\Hzero,\calO(t)}
\:.
\en
The hamiltonian of the perturbed system
is $H = \Hzero + \Delta H(t)$.
The perturbation, $\Delta H(t)$, vanishes at early times.
The perturbed system starts at early time unperturbed
and in equilibrium.
The density matrix, $\rho(t)$,
starts at early time equal to
the unperturbed equilibrium density
matrix, $\rhoequil$.
It evolves in time by
\eq
\partial_{t} \rho(t)
= - \frac{i}\hbar \comm{H,\rho(t)}
\en
The expectation values in the perturbed system,
$\expval{\calO(t)} = \Tr [\rho(t) \calO(t)]$,
evolve in time by
\eq
\partial_{t} \expval{\calO(t)}
=
\expval{ \partial_{t}\calO(t)}
+ \expval{ \frac{i}\hbar \comm{\Delta H(t),\calO(t)}}
\:.
\label{eq:time-evolution}
\en
This time evolution formula is especially useful when the 
equal-time commutator can be evaluated explicitly.

The entropy density evolves in time by
\eqa
\partial_{t} \expval{\rhoent(x,t)}
&=& \expval{ \partial_{t}\rhoent(x,t)}
+ \expval{ \frac{i}\hbar \comm{\Delta H(t),\rhoent(x,t)}} 
\nonumber\\
&=& \expval{ -\partial_{x}\Jent(x,t)}
+ \expval{ \frac{i}\hbar \comm{
\int \dif x' \,\,  \rhoent(x',t)
\Phient(x',t)
,\rhoent(x,t)}}
\ena
or
\eq
\partial_{t} \expval{\rhoent(x,t)}
+\partial_{x}\expval{ \Jent(x,t)}
=
\int \dif x' \,\, \Phient(x',t)
\expval{ \frac{i}\hbar \comm{
\rhoent(x',t)
,\rhoent(x,t)}}
\:.
\label{eq:continuity1}
\en
First, integrate over $x$
to find the rate of change of the total entropy:
\eqa
\frac{\dif S}{\dif t}
&=&
\int \dif x' \,\, \Phient(x',t)
\expval{ \frac{i}\hbar \comm{
\rhoent(x',t)
,\kB\beta \Hzero}} \nonumber\\
&=&
\kB\beta\int \dif x' \,\, \Phient(x',t)
\expval{ -\partial_{t}\rhoent(x',t)} \nonumber\\
&=&
\kB\beta\int \dif x' \,\, \Phient(x',t)
\expval{ \partial_{x}\Jent(x',t)} \nonumber\\
&=&
\kB\beta\int \dif x' \,\, \Eent(x',t)
\expval{ \Jent(x',t)}
\ena
which can be written
\eq
\frac{\dif S}{\dif t}
= \kB\beta \frac{\dif W}{\dif t}
\en
where $W(t)$
is the work done on the system by the external entropic
potential,
given by
\eq
\frac{\dif W}{\dif t}
=
\frac{\dif \expval{\Hzero}}{\dif t \hfill} 
=
\int \dif x \,\, \Eent(x,t)
\expval{ \Jent(x,t)}
\:.
\en
Equation~(\ref{eq:continuity1}),
unintegrated,
describes the local production of entropy:
\eq
\partial_{t} \expval{\rhoent(x,t)}
+\partial_{x}\expval{ \Jent(x,t)}
=
\kB \beta \, p(x,t)
\en
where
\eq
p(x,t) = 
\int \dif x' \,\, \Phient(x',t)
\expval{ \frac{i}\hbar \comm{\rhoent(x',t),T_{t}^{t}(x,t)}}
\label{eq:localpower}
\en
is the density of power delivered to the system
by the imposed entropic field.
The power delivered to the system by the external field
is a source of entropy.
This does not 
have an analogue in the 
continuity equation for electric charge.

The equal-time commutator appearing in 
(\ref{eq:continuity1}) and~(\ref{eq:localpower})
is calculated in 
Appendix~\ref{detail:equaltime}:
\eq
\frac{i}\hbar \comm{
T_{t}^{t}(x',t)
,T_{t}^{t}(x,t)}
=
\partial_{x'}\delta(x'-x) T_{t}^{x}(x,t)
-\partial_{x}\delta(x'-x) T_{t}^{x}(x',t)
\:.
\label{eq:equaltimeA}
\en
Substituting in (\ref{eq:continuity1}) gives
the continuity equation for entropy:
\eq
\begin{split}
\partial_{t} \expval{\rhoent(x,t)}
+\partial_{x}\expval{\Jent(x,t)}
&=
\kB \beta \Eent(x,t)  \expval{\Jent(x,t)}
\\
&\qquad
- 
\kB \beta \partial_{x} 
\left [
\Phient(x,t)
\expval{  \Jent(x,t)}
\right ]
\:.
\end{split}
\en
The form of this equation suggests 
that it might be worthwhile to redefine the entropy current
as $\Jent(x,t)+\kB \beta\Phient(x,t)\Jent(x,t)$
away from equilibrium,
so that the power will be proportional to the current.

The continuity equation for entropy is exact.
In the linear response approximation,
it becomes
\eq
\begin{split}
\partial_{t} \Delta\expval{\rhoent(x,t)}
+\partial_{x}\Delta\expval{ \Jent(x,t)}
&=
\kB \beta \Delta \Eent(x,t)  \expvalequil{\Jent(x,t)}
\\
&\qquad
- 
\partial_{x} 
\left [
\kB \beta \Delta \Phient(x,t)
\expvalequil{  \Jent(x,t)}
\right ]
\:.
\end{split}
\en
where, for any operator $\calO$,
\eq
\Delta\expval{\calO}=\expval{\calO}-\expvalequil{\calO}
\:.
\en
Entropy is locally conserved
to first order in the perturbation,
unless there is a persistent equilibrium
entropy current, $\expvalequil{\Jent(x,t)}$.
The distribution of entropy is stationary in equilibrium, 
$\partial_{t} \expvalequil{\rhoent(x,t)}=0$, so
$\partial_{x}\expvalequil{\Jent(x,t)}=0$.  
Therefore, if 
there is an
equilibrium entropy current, it must flow in closed loops within 
the circuit and must flow uniformly within each wire.
Since $\partial_{x}\expvalequil{\Jent(x,t)}=0$,
the linearized continuity equation can be written
\eq
\partial_{t} \Delta\expval{\rhoent(x,t)}
+\partial_{x}\Delta\expval{ \Jent(x,t)}
=
2\kB \beta \Delta \Eent(x,t) \expvalequil{\Jent(x,t)}
\:.
\label{eq:conslinearresp}
\en

Entropy current could be stored
in such persistent loops.
The possibility of
persistent entropy currents
in equilibrium 
is due to
the isolation of the near-critical degrees of 
freedom from the environment.
The operator that generates translations around 
a closed loop of wire $\mathcal{L}$  (disconnected from the rest of the 
circuit),
\eq
P_{\mathcal{L}} = \int_{\mathcal{L}} dx \; T^{t}_{x}(x,t) = (k\beta)^{-1} \int_{\mathcal{L}} dx \;\Jent(x,t)
\,,
\en
commutes with the hamiltonian.  Therefore the eigenspaces of 
$P_{\mathcal{L}}$ are superselection sectors, and there exists an equilibrium state for 
each eigenvalue of $P_{\mathcal{L}}$.  In the superselection sectors where $P_{\mathcal{L}}\ne 0$,
there is a nonzero persistent equilibrium
entropy current, $\expvalequil{\Jent(x,t)}\ne 0$.

\section{The entropy conduction equation}
The time evolution of the entropy current is
\eq
\partial_{t}\expval{\Jent(x,t)} =
\expval{\partial_{t}\Jent(x,t)}
+ 
\int \dif x' \,\, \Phient(x',t)
\expval{ \frac{i}\hbar \comm{
 \rhoent(x',t)
,\Jent(x,t)}}
\:.
\label{eq:timeev}
\en
Symmetry of the energy-momentum tensor implies
\eq
\Jent(x,t) = \kB\beta  T_{t}^{x}(x,t)
= - \kB\beta v^{2}  T_{x}^{t}(x,t)
\:.
\en
Local conservation of momentum implies
\eq
\partial_{t}\Jent(x,t) =
- \kB\beta v^{2}  \partial_{t}T_{x}^{t}(x,t)
= \kB\beta v^{2} \partial_{x} T_{x}^{x}(x,t)
\:.
\en
Equation~(\ref{eq:timeev}) becomes
\eq
\partial_{t}\expval{\Jent(x,t)} =
\kB\beta v^{2} \partial_{x}\expval{T_{x}^{x}(x,t)}
+ \int \dif x' \,\, \Phient(x',t)
\expval{ \frac{i}\hbar \comm{
\rhoent(x',t)
,\Jent(x,t)}}
\:.
\label{eq:timeevolA}
\en
The equal-time commutator appearing in 
(\ref{eq:timeevolA}) is calculated in 
Appendix~\ref{detail:equaltime}:
\eq
\begin{split}
\frac{i}\hbar \comm{T_{t}^{t}(x',t), v^{-2} T_{t}^{x}(x,t)}
&=
\frac{\cUV}{6} \frac{\hbar v}{2\pi} \partial_{x}^{3} \delta(x'-x)
\\
&\qquad
{}
+\partial_{x}\delta(x'-x)
[T_{x}^{x}(x,t)-T_{t}^{t}(x,t)]
\\
&\qquad
{}
+\delta(x'-x)  \partial_{x}T_{x}^{x}(x,t)
\label{eq:equaltimeB}
\end{split}
\en
where $\cUV$ is the bulk conformal central charge in the 
short-distance limit. 
Substituting in (\ref{eq:timeevolA}) gives
the conduction equation:
\eq
\begin{split}
\partial_{t}\expval{\Jent(x,t)} &=
- 
\kB^{2} \beta^{2} v^{2} 
\frac{\cUV}{6} \frac{\hbar v}{2\pi} \partial_{x}^{2}\Eent(x,t)
\\
&\qquad{}
+
\kB^{2} \beta^{2} v^{2}
\expval{T_{t}^{t}(x,t)-T_{x}^{x}(x,t)}
\Eent(x,t)
\\
& \qquad 
{}
+
\left [
1
+\kB \beta \Phient(x,t)
\right ]
\kB\beta v^{2}
\partial_{x}\expval{T_{x}^{x}(x,t)}
\:.
\label{eq:timeevolfinal}
\end{split}
\en
The conduction equation is especially useful in
two situations:
when the entropy flow in the wire is uniform,
so $\partial_{x}\expval{T_{x}^{x}(x,t)}=0$,
or when the wire is conformally invariant in the bulk,
so $T_{x}^{x}=-T^{t}_{t}$.

\section{Uniform entropy conduction}
When the entropy flow is uniform, the spatial derivatives
in (\ref{eq:timeevolfinal}) are negligible,
so the conduction equation becomes
\eq
\partial_{t}\expval{\Jent(x,t)} =
\kB^{2} \beta^{2} v^{2} 
\expval{T_{t}^{t}(x,t)-T_{x}^{x}(x,t)}
\Eent(x,t)
\:.
\en
In the linear response approximation,
this becomes
\eq
\partial_{t}\Delta \expval{\Jent(x,t)} =
\kB^{2} \beta^{2} v^{2} 
\expvalequil{T_{t}^{t}(x,t)-T_{x}^{x}(x,t)}
\Delta \Eent(x,t)
\:.
\label{eq:timeevoluniform}
\en
The quantity
$\kB \beta \expvalequil{T_{t}^{t}(x,t)-T_{x}^{x}(x,t)}$
is the equilibrium entropy density, $\Sdensity$,
by the following argument.\cite{Zamolodchikov:1989cf}
The equilibrium energy density
in a long wire of length $l$ is
\eq
\Edensity = \expvalequil{T_{t}^{t}(x,t)} =
-\partialby\beta \frac{\ln Z_{0}}{l}
\:.
\en
The free energy density in the one-dimensional volume $V=l$,
\eq
\Fdensity = - \frac1\beta \frac{\ln Z_{0}}{V} = - \frac1\beta \frac{\ln Z_{0}}{l}
\:,
\en
is independent of $l$ in the limit 
of large $l$, so
\eq
\Fdensity = (1+l \partialby{l}) \Fdensity
=  - \frac1\beta \partialby{l} \ln Z_{0} 
= \expvalequil{T_{x}^{x}(x,t)}
\label{eq:Txx}
\:,
\en
which is the standard thermodynamic relation
between the specific free energy and the pressure $P = kT d\ln Z/dV$.
The equilibrium entropy density is
\eq
\Sdensity = 
\kB \left (1- \beta \partialby\beta \right )\frac{\ln Z_{0}}{l}
= \kB \beta (\Edensity-\Fdensity)
= \kB \beta^{2} \partialby \beta \Fdensity
\:.
\en
It can now be written in two equivalent ways:
\eq
\Sdensity
= \kB \beta \expvalequil{T_{t}^{t}(x,t)-T_{x}^{x}(x,t)}
=\kB \beta^{2} \partialby \beta \expvalequil{T_{x}^{x}(x,t)}
\label{eq:Sdensity}
\:.
\en
Substituting in the formula for uniform conduction, 
(\ref{eq:timeevoluniform}), gives
\eq
\partial_{t}\Delta \expval{\Jent(x,t)} =
\kB \beta v^{2} \Sdensity \Delta \Eent(x,t)
\:.
\en
The entropic conductivity
for uniform entropy flow is therefore
\eq
\sigmaent(\omega)= \frac{i \kB \beta v^{2} 
\Sdensity}\omega
= \frac{i v^{2} \Sdensity}{\omega T}
\:.
\label{eq:sigmaentgeneral}
\en
The complex thermal conductivity is
\eq
\sigma_{\mathit{thermal}}(\omega) = T  \sigmaent(\omega)
=
\frac{i v^{2} \Sdensity}{\omega}
\en
and the thermal conductivity is
\eq
\kappa(\omega,T)  = \Real( T \sigmaent(\omega)) =
\pi v^{2} \Sdensity \,\delta(\omega)
\:.
\en
An alternative derivation of 
(\ref{eq:sigmaentgeneral}) is given in 
Appendix~\ref{app:kubo}, using the Kubo formula for the 
conductivity instead of the linearized conduction equation.

As a consistency check, take the static limit of
the conduction formula, (\ref{eq:timeevolfinal}),
make the linear response approximation,
then take the uniform limit,
letting the entropic potential become constant in space,
$\Delta \Phient(x,t)=-\Delta T$.
The perturbed system will be in equilibrium at temperature
$T+\Delta T$.
The conduction equation becomes
\eq
0= \kB \beta v^{2} \Delta\expval{T_{x}^{x}(x,t)}
+\kB^{2} \beta^{2} v^{2} \expvalequil{T_{t}^{t}(x,t)-T_{x}^{x}(x,t)} \Delta T
\en
or
\eq
0= \Delta \Fdensity + \Sdensity \Delta T 
\en
which is
just the thermodynamic relation
between free energy and entropy.

\section{Bulk-critical wire}
\label{sect:bulkcriticalwire}
When the quantum wire is critical in the bulk,
the 1+1 dimensional quantum field theory
is conformally invariant in the bulk:
\eq
\Theta(x,t) = - T_{\mu}^{\mu}(x,t)
= - T_{t}^{t}(x,t)- T_{x}^{x}(x,t)  = 0
\:.
\en
Then $T_{x}^{x}(x,t)$ can be replaced
by $-T_{t}^{t}(x,t)$ in
the conduction equation, (\ref{eq:timeevolfinal}), giving
\eqa
\lefteqn{
\partial_{t}\expval{\Jent(x,t)} =
-\kB\beta v^{2} \partial_{x}\expval{T_{t}^{t}(x,t)}
- 
\kB^{2} \beta^{2} v^{2} 
\frac{c}{6} \frac{\hbar v}{2\pi} \partial_{x}^{2}\Eent(x,t)
}\hspace{2em} \nonumber \\[1ex]
&&
{}+
\kB^{2} \beta^{2} v^{2} \Eent(x,t)
2 \expval{T_{t}^{t}(x,t)}
-\kB^{2} \beta^{2} v^{2} \Phient(x,t)
\partial_{x}\expval{T_{t}^{t}(x,t)}
\:.
\ena
In the linear response approximation, this becomes
\eq
\begin{split}
\partial_{t}\Delta\expval{\Jent(x,t)}
&+ v^{2} \partial_{x}\Delta\expval{\rhoent(x,t)}
=
\\
&
\kB^{2} v^{2} \beta^{2}
\left [
2 \expvalequil{T_{t}^{t}(x,t)}
- 
\frac{c}{6} \frac{\hbar v}{2\pi} \partial_{x}^{2}
\right ] \Delta \Eent(x,t)
\:.
\end{split}
\en
For bulk-critical wire, the equilibrium energy density 
is\cite{BLOTECARDYNIGHTINGALE,AFFLECK1986}
\eq
\Edensitycritical = \expvalequil{T_{t}^{t}(x,t)}
=\frac{c}{12} \frac{2\pi}{\hbar v} \frac1{\beta^{2}}
\en
so the linearized conduction equation is
\eq
\begin{split}
\partial_{t}\Delta\expval{\Jent(x,t)}
+ v^{2} \partial_{x}\Delta \expval{\rhoent(&x,t)}
=
\\
&
\kB^{2}  \frac{c}{6}
\frac{2\pi v}{\hbar}
\left [ 1 - \left (\frac{\hbar v}{2\pi}\right )^{2}
\beta^{2}  \partial_{x}^{2}
\right ]
\Delta \Eent(x,t)
\:.
\end{split}
\en
This equation and
the linearized continuity equation
(\ref{eq:conslinearresp})
together determine the entropy flow in the wire.
For 
$
\Delta \Eent(x,t) = \me^{iqx-i\omega t+\epsilon t} \Delta \Eent(0,0)
$,
the entropy current is
\eq
\Delta \expval{\Jent(x,t)}
= \sigmaent(q,\omega) \Delta \Eent(x,t)
\:.
\en
with
\eqa
\lefteqn{\sigmaent(q,\omega) =
\kB^{2} \frac{2\pi v}{\hbar} \frac{c}{12}
\left [ 1
+ \left ( \frac{\hbar v\beta q}{2\pi} \right ) ^{2}\right ]
\left (
\frac{i}{\omega+i\epsilon-vq}
+\frac{i}{\omega+i\epsilon+vq}
\right ) }\hspace{9em}
\nonumber \\
&&{} + v\kB \beta \expvalequil{\Jent}
\left (
\frac{i}{\omega+i\epsilon-vq}
-\frac{i}{\omega+i\epsilon+vq}
\right )
.
\label{eq:sigmaentqomegacritical}
\ena
The change in the entropy density is
\eq
\begin{split}
\Delta \expval{\rhoent(x,t)}
&=
\kB^{2} \frac{2\pi}{\hbar} \frac{c}{12}
\left [ 1 + \left ( \frac{\hbar v\beta q}{2\pi} \right ) ^{2}\right ]
\\
&\qquad\qquad\qquad
\left (\frac{i}{\omega+i\epsilon-vq}-\frac{i}{\omega+i\epsilon+vq}
\right )
\Delta \Eent(x,t)
\\
& \qquad{} + \kB \beta \expvalequil{\Jent}
\left (
\frac{i}{\omega+i\epsilon-vq}
+\frac{i}{\omega+i\epsilon+vq}
\right )
\Delta \Eent(x,t)
.
\label{eq:inducedentropy}
\end{split}
\en
In the limit of uniform flow,
\eq
\sigmaent(\omega) =
\lim_{q \rightarrow 0}
\sigmaent(q,\omega) =
\frac{c}{6}
\frac{2\pi \kB^{2}  v}{\hbar} 
\frac{i }\omega
\label{eq:sigmaentcritical}
\en
which
agrees with the general formula for the entropic conductivity, 
(\ref{eq:sigmaentgeneral}),
since
the equilibrium entropy density
of bulk-critical wire is
\eq
\Sdensitycritical = \frac{c}{6} \frac{2\pi}{\hbar v} 
\frac{\kB}{\beta}
\label{eq:Sdensitycritical}
\:.
\en
Equation~(\ref{eq:sigmaentcritical}) is equivalent to the
formula for the thermal conductivity of 
bulk-critical quantum wire,
 $\kappa(\omega,T) = (\kB^{2}\pi^{2} T v c/3\hbar ) 
\delta(\omega) $,
which was previously derived in \onlinecite{orignac:134426}.

As another check,
take the static limit, $\Delta \Eent(x,t)=-\partial_{x}\Delta \Phient(x)$.
Equation (\ref{eq:inducedentropy}) gives the change in the entropy 
density:
\eq
\Delta\expval{\rhoent(x,t)}
= 
- \kB^{2} \frac{2\pi}{\hbar v} \frac{c}{6}
\left [ 1
- \left ( \frac{\hbar v\beta }{2\pi} \right )^{2}
\partial_{x}^{2}\right ]
\Delta \Phient(x)
\:.
\label{eq:entropychange}
\en
Let the entropic potential become uniform,
$\Delta \Phient(x)=-\Delta T$.
The system should respond to the perturbation by
going to equilibrium at temperature $T+\Delta T$.
The induced change in entropy density will be,
according to (\ref{eq:entropychange}),
\eq
\Delta\expval{\rhoent(x,t)}
= 
\kB^{2} \frac{2\pi}{\hbar v} \frac{c}{6}
\Delta T
\en
which agrees with the temperature derivative of
the equilibrium entropy density,
(\ref{eq:Sdensitycritical}).

\section{Chiral energy and entropy currents}
\label{sect:chiralcurrents}
The energy-momentum tensor has only two independent 
components
when the quantum wire is critical in the bulk.
They can be written as two currents
\eqa
\TR(x,t)&=& \frac12 (v T_{t}^{t}(x,t) +  T_{t}^{x}(x,t))
\\[1ex]
\TL(x,t)&=&  \frac12 (v T_{t}^{t}(x,t) - T_{t}^{x}(x,t))
\:.
\ena
The conservation laws,
$\partial_{\mu}T^{\mu}_{\nu}(x,t)=0$, become
\eq
(\partial_{t}+v\partial_{x})\TR(x,t) =
(\partial_{t}- v\partial_{x})\TL(x,t) = 0 
\:,
\en
so each is a chiral current, depending on a single 
coordinate:
\eqa
\TR(x,t)&=& \TR(x-vt)
\label{eq:TRB}\\
\TL(x,t) &=&\TL(x+vt)
\label{eq:TLB}
\:.
\ena
The entropy current is a sum of chiral entropy currents:
\eqa
\Jent(x,t) &=& \JR(x,t) - \JL(x,t) \\
\rhoent(x,t) &=& \frac1v\JR(x,t) + \frac1v\JL(x,t) 
\ena
where
\eqa
\JR(x,t)&=&\JR(x-vt) 
=\kB \beta \TR(x,t) -\frac12 \kB \beta v\expvalequil{T_{t}^{t}(x,t)}
\\
\JL(x,t) &=& \JL(x+vt)
= \kB \beta \TL(x,t)-\frac12 \kB \beta v\expvalequil{T_{t}^{t}(x,t)}
\:.
\ena
$\JR(x,t)$ flows to the right,
$\JR(x,t) = \JR(x+v\delta t,t+\delta t)$,
and $\JL(x,t)$ flows to the left,
$\JL(x,t) = \JL(x-v\delta t,t+\delta t)$,
both at the speed of ``light'', $v$.

The chiral energy currents are, up to normalization, the 
usual chiral components, $T(z)$ and $\bar T(\bar z)$, of the 
euclidean energy-momentum tensor: $\TR(z) = -\hbar v^{2} 
T(z)/2\pi$, $\TL(\bar z) = -\hbar v^{2} \bar T(\bar 
z)/2\pi$, where $z= x+iv\tau$, $\bar z = x-iv\tau$, 
$\tau=it$.  The usual analytic operator product expansions 
of $T(z)$ and $\bar T(\bar z)$ are equivalent 
to equal-time commutation relations
(see Appendix~B of \onlinecite{DFEntropyFlowII} (Appendix~2 of 
\onlinecite{DFEntropyFlowIIJStatPhys})
for details).
These are equivalent to the general equal-time commutation 
relations
derived in Appendix~\ref{detail:equaltime},
specialized to the bulk-critical case.

A naive calculation shows the condition 
for uniform entropy flow in bulk-critical wire.
Consider a wire of 
length $l$. 
Let $I(t)=I(0)\cos(i\omega t)$ be
the entropy current
flowing into the wire 
from the left, at $x=-l/2$.
The same entropy current flows out 
of the wire to the right, at $x=+l/2$.
The boundary conditions
completely determine the expectation values of 
the chiral currents,
\eqa
\expval{\Jent(-\frac{l}2 ,t)}
&=& \expval{\JL(-\frac{l}2-vt)}-\expval{\JR(-\frac{l}2+vt)}
=I(0)\cos(i\omega t) \\
\expval{\Jent(+\frac{l}2,t)}
&=&\expval{\JL(+\frac{l}2-vt)}-\expval{\JR(+\frac{l}2+vt)}
=I(0)\cos(i\omega t)
\:,
\ena
so the entropy
current inside the wire is
\eq
\expval{\Jent(x,t)} = I(t) \frac{\cos(\omega x/v)}{\cos(\omega l/2v)}
\:.
\en
The condition for uniform flow
is therefore $\omega l\ll v$.

\section{Conclusion}

An argument has been presented that near-critical quantum 
circuits are, in principle, ideal physical systems for 
large-scale quantum computers,
because they are effectively isolated and controllable.
The relativistic quantum 
field theories in 1+1 dimensions are then universal 
``machine languages'' for large-scale quantum circuit 
computers.  It was remarked that laws governing the flow of 
entropy are basic constraints on the design of reversible 
quantum computers, and that entropy flows in near-critical 
quantum circuits as a conserved quantum current, so circuit 
laws can be written for entropy flow in analogy with the 
electric circuit laws.

It was argued that the quantum wires should be stably 
bulk-critical, with no relevant bulk couplings, to avoid 
intractable control problems in the bulk wires.  It was 
pointed out that bulk-critical quantum wires have some 
possibly useful features: all excitations move along the 
wires at a uniform speed, $v$, and the entropy current 
separates into left and right moving chiral currents which 
do not interact with each other in the bulk.

The continuity equation for entropy, 
(\ref{eq:continuity}), and the equation for the 
conduction of entropy in wires, (\ref{eq:conduction}), 
were derived.  The entropy continuity and conduction 
equations are exact.  Neither depends on a linear response 
approximation.  They are universal equations, because they 
follow from universal equal-time commutation relations of 
the energy density and current.

The conduction equation, in the limit of uniform flow, was 
shown to imply a formula for the entropic conductivity of 
near-critical quantum wire, $\sigmaent(\omega) = iv^{2} 
\Sdensity /\omega T $, equivalent to a formula for the 
thermal conductivity, $\kappa(\omega,T) = \pi v^{2} 
\Sdensity \delta(\omega)$.  This formula provides a way to 
measure directly, by experiment, the entropy density, 
$\Sdensity$, of the low-energy excitations in 
one-dimensional and quasi-one-dimensional near-critical 
quantum systems.

It is hoped that these will be useful preliminary steps 
towards the design of near-critical quantum circuits that 
can perform large-scale quantum computations.

\vskip2ex

\noindent
{\bf Acknowledgements}\\
I thank A.~Konechny for many discussions.
I thank
the members of an informal Rutgers seminar ---
S.~Ashok, A.~Ayyer, D.~Belov, E. Dell'Aquila,
B.~Doyon, and R.~Essig
--- for listening to a preliminary version of this work,
and for their comments and questions.
I thank
M.~Douglas and G.~Moore for reminding me that the 
monster conformal field theory is an example of a 
completely stable renormalization group fixed point in 1+1 
dimensions,
and G.~Moore for pointing out \onlinecite{dixon:1988}.
I thank
S.~Lukyanov for pointing towards some of the
condensed matter literature, leading in particular to
\onlinecite{luttinger:1964,orignac:134426}.
I thank
N.~Andrei for helpful comments on the manuscript
and for explaining to me that there are 
quantum critical phenomena which are
not described by relativistic 
quantum field theories.
This work was supported by the Rutgers New High Energy 
Theory Center.

\newpage
\appendix
\addtocontents{toc}{\protect\setcounter{tocdepth}{0}}
\renewcommand{\thesection}{Appendix \arabic{section}:}
\section{Equal-time commutators of $T_{t}^{t}(x,t)$
and $T_{x}^{t}(x,t)$}
\label{detail:equaltime}
The universal equal-time commutators of the energy and momentum
densities
\eqa
\frac{i}{\hbar} \comm{T_{t}^{t}(x',t), \,T_{t}^{t}(x,t)}
&=&
-\partial_{x}\delta(x'-x) 2 T_{t}^{x}(x,t)
- \delta(x'-x) \partial_{x} T_{t}^{x}(x,t)
\label{equaltimeTTA}\\
\frac{i}{\hbar} \comm{T_{x}^{t}(x',t), \,T_{x}^{t}(x,t)}
&=&
\partial_{x}\delta(x'-x) 2 T_{x}^{t}(x,t)
+ \delta(x'-x) \partial_{x} T_{x}^{t}(x,t)
\label{equaltimeTTB}\\
\frac{i}{\hbar} \comm{T_{t}^{t}(x',t), \,T_{x}^{t}(x,t)}
&=&
- \frac\cUV{6} \frac{\hbar v}{2\pi} \partial_{x}^{3} \delta(x'-x)
+\partial_{x}\delta(x'-x) [T_{t}^{t}(x,t)-T_{x}^{x}(x,t)]
\nonumber \\
&&\qquad\qquad
{}
-\delta(x'-x)\partial_{x}T_{x}^{x}(x,t)
\label{equaltimeTTC}
\ena
are derived here
from the Ward identities for the operator product of two 
energy-momentum tensors.
The number $\cUV$ is the bulk conformal central charge at 
short-distance.

Make an infinitesimal local variation of the space-time metric,
$g_{\mu\nu}\rightarrow g_{\mu\nu} + \delta g_{\mu\nu}(x,t)$,
combined with an infinitesimal space-time transformation,
$x^{\mu}\rightarrow x^{\mu}+\delta x^{\mu}(x,t)$.
The combined change in the metric is
\eq
g_{\mu\nu}\rightarrow g_{\mu\nu} +\partial_{\mu}(\delta x_{\nu})
+\partial_{\nu}(\delta x_{\mu}) 
+ \delta g_{\mu\nu}
+ \delta x^{\alpha}\partial_{\alpha}(\delta g_{\mu\nu})
+ \partial_{\mu}(\delta x^{\alpha})\delta g_{\alpha\nu}
+ \partial_{\nu}(\delta x^{\alpha})\delta g_{\mu\alpha}
\:.
\en
Vary $\ln Z$, keeping terms that are
first order in $\delta x^{\mu}$
and in $\delta g_{\mu\nu}$,
to obtain the Ward identity
on the time-ordered product of two energy-momentum tensors:
\eqa
\frac{i}\hbar \,\partial'_{\mu'} \,
t\{
T_{\nu'}^{\mu'}(x',t') \,T_{\nu}^{\mu}(x,t) 
\}
&=& \partial_{\alpha}\left [ \delta(x'-x)\delta(t'-t)\right ] 
\left ( 
\delta^{\alpha}_{\nu'} T_{\nu}^{\mu}
-g_{\nu'\nu}g^{\alpha\beta}T_{\beta}^{\mu}
-\delta^{\mu}_{\nu'} T^{\alpha}_{\nu}
\right ) (x,t)
\nonumber \\
&&\qquad\qquad{}+
\delta(x'-x)\delta(t'-t) \partial_{\nu'}T_{\nu}^{\mu}(x,t)
\:.
\ena
Integrate both sides of the Ward identity 
over $t'$ from $t-\epsilon$ to $t+\epsilon$
to obtain:
\eqa
\lefteqn{
\frac{i}\hbar \comm{T_{\nu'}^{t}(x',t), \,T_{\nu}^{\mu}(x,t) }
+
\partial_{x'}\,
\int_{t-\epsilon}^{t+\epsilon} \dif t' \,\,
\frac{i}\hbar\,
t\{
T_{\nu'}^{x}(x',t') \,T_{\nu}^{\mu}(x,t) 
\} =}
\nonumber \\
&&\int_{t-\epsilon}^{t+\epsilon} \dif t' \,\,
\bigg \{
\partial_{\alpha}\left [ \delta(x'-x)\delta(t'-t)\right ]
 \left ( 
\delta^{\alpha}_{\nu'} T_{\nu}^{\mu}
-g_{\nu'\nu}g^{\alpha\beta}T_{\beta}^{\mu}
-\delta^{\mu}_{\nu'} T^{\alpha}_{\nu}
\right ) (x,t) 
\nonumber \\
&& \qquad \qquad \qquad {}+
\delta(x'-x)\delta(t'-t) \partial_{\nu'}T_{\nu}^{\mu}(x,t)
\bigg \}
\label{eq:equaltimeI}
\:.
\ena
The time integral on the lhs picks out the contact terms in the 
time-ordered operator product.
The energy-momentum tensor has scaling dimension 2,
so the contribution of the contact terms has the form:
\eqa
\lefteqn{
\int_{t-\epsilon}^{t+\epsilon} \dif t' \,\,
\frac{i}\hbar\,
t\{
T_{\nu'}^{\mu'}(x',t') \,T_{\nu}^{\mu}(x,t) 
\}
=
}
\nonumber \\
&&
\int_{t-\epsilon}^{t+\epsilon} \dif t' \,\,
 \left [
C_{\nu'\nu}^{\mu'\mu\alpha\beta} \partial_{\alpha}\partial_{\beta}
+B_{\nu'\nu}^{\mu'\mu\alpha}(x,t) \partial_{\alpha}
+ A_{\nu'\nu}^{\mu'\mu}(x,t)
\right ]  \left [ \delta(x'-x)\delta(t'-t)\right ]
\label{eq:equaltimecontact}
\ena
for some operator-valued tensors $A$, $B$, $C$.
The operators $C_{\nu'\nu}^{\mu'\mu\alpha\beta}$ have scaling 
dimension 0, so are multiples of the identity.

By (\ref{eq:equaltimeI})
and~(\ref{eq:equaltimecontact}), the equal-time commutators of 
the energy and momentum densities are:
\eqa
\frac{i}{\hbar} \comm{T^{t}_{\nu'}(x',t), \,T^{t}_{\nu}(x,t)}
&=&
c_{\nu'\nu}\partial_{x}^{3}\delta(x'-x)
 +
b_{\nu'\nu}(x,t) \partial_{x}^{2}\delta(x'-x)
\nonumber \\
&&
{}+ \left (
\delta_{\nu'}^{x} T_{\nu}^{t}
- \delta_{\nu'}^{t} T_{\nu}^{x}
-g_{\nu'\nu} T_{x}^{t}
+a_{\nu'\nu}
\right ) (x,t)
\partial_{x}\delta(x'-x)
\nonumber \\
&&{}+\partial_{\nu'}T_{\nu}^{t}(x,t) \delta(x'-x)
\ena
where $a_{\nu'\nu}=A^{tt}_{\nu'\nu}$,
$b_{\nu'\nu}=B_{\nu'\nu}^{ttxx}$,
and $c_{\nu'\nu}=C_{\nu'\nu}^{ttxx}$.

The antisymmetry of the commutators is equivalent to:
\eqa
0 &=& c_{\nu'\nu}-c_{\nu\nu'}\\
0 &=& b_{\nu'\nu}+b_{\nu\nu'} \\
0 &=& \partial_{x} (a_{xx}-2T^{t}_{x}) \\
0 &=& \partial_{x} a_{tt} \\
0 &=& \partial_{x} (a_{xt}+a_{tx}+T^{x}_{x}-T^{t}_{t}) \\
0 &=& a_{tx}-a_{xt} - 2 \partial_{x}b_{tx}
-T_{t}^{t} - T_{x}^{x}
\:.
\ena
Therefore $b_{xx}=b_{tt}=0$, $b_{xt}=-b_{tx}$,
and, up to multiples of the identity operator,
\eqa
a_{xx}&=&2T^{t}_{x} \\
a_{tt}&=&0 \\
a_{xt} &=& \partial_{x}b_{tx}- T^{x}_{x} \\
a_{tx} &=& \partial_{x}b_{tx} + T^{t}_{t}
\:.
\ena
Ignoring multiples of the identity operator for the time 
being, the only unknown is the operator $b_{tx}(x,t)$.
The equal-time commutators are,
up to multiples of the identity,
\eqa
\frac{i}{\hbar} \comm{T_{t}^{t}(x',t), \,T_{t}^{t}(x,t)}
&=&
-\partial_{x}\delta(x'-x) 2 T_{t}^{x}(x,t)
- \delta(x'-x) \partial_{x} T_{t}^{x}(x,t)
\label{equaltimeTTAtemp}\\
\frac{i}{\hbar} \comm{T_{x}^{t}(x',t), \,T_{x}^{t}(x,t)}
&=&
\partial_{x}\delta(x'-x) 2 T_{x}^{t}(x,t)
+ \delta(x'-x) \partial_{x} T_{x}^{t}(x,t)
\label{equaltimeTTBtemp}\\
\frac{i}{\hbar} \comm{T_{t}^{t}(x',t), \,T_{x}^{t}(x,t)}
&=&
+\partial_{x}\delta(x'-x) (T_{t}^{t}-T_{x}^{x})(x,t)
-\delta(x'-x)\partial_{x}T_{x}^{x}(x,t)
\nonumber \\
&&\qquad\qquad
{} + \partial_{x} \left [
\partial_{x}\delta(x'-x) b_{tx}(x,t)
\right ]
\label{equaltimeTTCtemp}
\:.
\ena
Take the time derivative of both sides
of (\ref{equaltimeTTAtemp}).
In the time derivative of (\ref{equaltimeTTAtemp}),
use (\ref{equaltimeTTCtemp}) to evaluate the 
commutators.
The equation that results is:
\eq
0 = 2 \partial_{x}^{3}\delta(x'-x)b_{tx}(x,t)
+3 \partial_{x}^{2}\delta(x'-x)\partial_{x}b_{tx}(x,t)
+\partial_{x}\delta(x'-x)\partial_{x}^{2}b_{tx}(x,t)
\:.
\en
So $b_{tx}(x,t)=0$.

Equations (\ref{equaltimeTTAtemp}--\ref{equaltimeTTCtemp}),
with $b_{tx}(x,t)=0$,
give 
the equal-time commutators up to multiples of the identity.
These are exactly
(\ref{equaltimeTTA}--\ref{equaltimeTTC}),
up to multiples of the identity.
So all that remains is to determine the
multiples of the identity operator that appear in the
equal-time commutators.

The terms proportional
to the identity operator in
(\ref{equaltimeTTA}--\ref{equaltimeTTC})
are determined
by evaluating the expectation values
of the equal-time 
commutators
in the ground-state.
The spectral representation
of the ground-state
two-point function of the energy-momentum tensor 
is:\cite{CFL}
\eq
\vev{
\frac{i}\hbar\,
t \{
T_{\nu'}^{\mu'}(x',t')
\,
T_{\nu}^{\mu}(x,t)
\}
}
=
\int_{0}^{\infty} \dif (m^{2})\,\rho_{c}(m^{2})
\;
G^{\mu'\mu}_{\nu'\nu}(x'-x,t'-t;m^{2})
\label{eq:spectral}
\en
\eq
G^{\mu'\mu}_{\nu'\nu}(x,t;\mu) =
\frac1{(2\pi)^{2}}
\int\int \dif p_{x}\dif p_{t} \,\,
\me^{i p_{x}x +i p_{t}t}
\frac{(p_{\nu'}p^{\mu'}-\delta_{\nu'}^{\mu'}p^{2})
(p_{\nu}p^{\mu}-\delta_{\nu}^{\mu}p^{2})}
{p_{\mu}p^{\mu}+m^{2}+i\epsilon}
\:.
\en
The conformal central charge in the short distance limit, $\cUV$,
is given by
\eq
\int \dif (m^{2})\,\rho_{c}(m^{2}) =\frac\cUV{6}
\frac{\hbar v}{2\pi}
\:.
\en
Extract the equal-time commutator from
(\ref{eq:spectral}) by evaluating
at $t'=t+\epsilon$ and at $t'=t-\epsilon$
and taking the difference:
\eqa
\lefteqn{
\vev{\frac{i}\hbar
\,\comm{T_{\nu'}^{\mu'}(x',t)\,T_{\nu}^{\mu}(x,t)}}
=
\int_{0}^{\infty} \dif (m^{2})\,\rho_{c}(m^{2})
\,\,\frac1{2\pi} \int \dif p_{x} \,\, \me^{i p_{x}(x'-x)}
}
\nonumber \\
&&\qquad\qquad \frac1{2\pi}
\int \dif p_{t} \,\,
\frac{\me^{ip_{t}\epsilon} - \me^{-ip_{t}\epsilon}}
{p_{\mu}p^{\mu}+m^{2}+i\epsilon}
\left [ (p_{\nu'}p^{\mu'}-\delta_{\nu'}^{\mu'}p^{2})
(p_{\nu}p^{\mu}-\delta_{\nu}^{\mu}p^{2})
\right ]
\:.
\ena
In particular,
\eqa
\vev{\frac{i}\hbar\comm{T_{t}^{t}(x',t)\,T_{t}^{t}(x,t)}}
&=& 0  \\
\vev{\frac{i}\hbar\comm{T_{x}^{t}(x',t)\,T_{x}^{t}(x,t)}}
&=& 0 \\
\vev{\frac{i}\hbar\comm{T_{t}^{t}(x',t)\,T_{x}^{t}(x,t)}}
&=&
- \partial_{x}^{3} \delta(x'-x)
\frac\cUV{6}
\frac{\hbar v}{2\pi}
\:.
\ena
This fixes the terms proportional
to the identity operator in
(\ref{equaltimeTTA}--\ref{equaltimeTTC}),
finishing their derivation.

\section{$\sigmaent(\omega)= iv^{2} \Sdensity 
/\omega T $ from the Kubo formula}
\label{app:kubo}

The Kubo formula for the entropy current induced in a wire 
by an entropic potential $\Delta \Vent(x,t)$ is
\eqa
\Delta \expval{\Jent(x_{2},t_{2})}
&=& 
\int_{-\infty}^{t_{2}}\dif t_{1} \,\,
\expvalequil{\frac{i}\hbar
\comm{\Delta H(t_{1}), \Jent(x_{2},t_{2})}} \nonumber \\
&=& 
\int_{-\infty}^{t_{2}}\dif t_{1} \,\,
\expvalequil{\frac{i}\hbar
\comm{\int\dif x_{1}\, \Delta \Vent(x_{1},t_{1}) 
\rhoent(x_{1},t_{1}), \Jent(x_{2},t_{2})}}
\:.
\ena
The Kubo formula is the solution of the
time evolution equation, (\ref{eq:time-evolution}),
in the linear response approximation.

For an alternating entropic potential,
$
\Delta \Vent(x,t)=\me^{iqx-i\omega t}\Delta \Vent(0,0)
$,
the induced current is
\eq
\Delta \expval{\Jent(x,t)}
=\sigmaent(q,\omega) \Delta \Eent(x,t)
\en
where $\Delta \Eent(x,t) = -iq \Delta \Vent(x,t)$.
The Kubo formula for the entropic conductivity is
\eqa
\sigmaent(q,\omega)
&=& 
\frac{i}{q}
\int\dif x_{1}\,
\int_{-\infty}^{t_{2}}\dif t_{1} \,\,
\me^{i\omega(t_{2}- t_{1})-iq(x_{2}-x_{1})}
\expvalequil{\frac{i}\hbar
\comm{ \rhoent(x_{1},t_{1}), \Jent(x_{2},t_{2})}} \nonumber \\
&=& 
\kB^{2}\beta^{2}
\frac{i}{q}
\int\dif x_{1}\,
\int_{-\infty}^{t_{2}}\dif t_{1} \,\,
\me^{i\omega(t_{2}- t_{1})-iq(x_{2}-x_{1})}
\\
\nonumber
&&\qquad\qquad\qquad\qquad\qquad
\expvalequil{\frac{i}\hbar
\comm{ T_{t}^{t}(x_{1},t_{1}), T_{t}^{x}(x_{2},t_{2})}}
\:.
\ena
Introduce the Fourier transform of the energy-momentum tensor:
\eq
\tilde{T}_{\nu}^{\mu}(p,\eta)
= \int \dif x \int \dif t \; \me^{i(\eta t-p x)}
T_{\nu}^{\mu}(x,t)
\:.
\en
Write its two-point functions:
\eq
\expvalequil{
\tilde{T}_{\nu'}^{\mu'}(p',\eta')
\, \tilde{T}_{\nu}^{\mu}(p,\eta)
}
= (2\pi)^{2} \delta(p'+p) \delta(\eta'+\eta)
G_{\nu'\nu}^{\mu'\mu}(p,\eta)
\:.
\en
The equilibrium
expectation values of the commutators are given by
\eq
\expvalequil{
\frac{i}\hbar
\comm{ 
\tilde{T}_{\nu'}^{\mu'}(p',\eta')
\, \tilde{T}_{\nu}^{\mu}(p,\eta)}
}
= (2\pi)^{2} \delta(p'+p) \delta(\eta'+\eta)
\frac{i}\hbar
\left (1-\me^{\beta\hbar \eta} \right )
G_{\nu'\nu}^{\mu'\mu}(p,\eta)
\:.
\en
The Kubo formula becomes
\eq
\sigmaent(q,\omega)
=
\frac{\kB^{2}\beta^{2}}{\hbar}
\int \dif \eta \;
\frac{1}{\omega+i\epsilon-\eta}
\left (1-\me^{\beta\hbar \eta} \right )
\frac1{iq}
G_{tt}^{tx}(q,\eta)
\:.
\en
Conservation and symmetry of the energy-momentum tensor imply
\eq
\eta \tilde{T}_{t}^{x}(q,\eta)
= -v^{2} q \tilde{T}_{x}^{x}(q,\eta)
\en
so
\eq
\frac1{q}
G_{tt}^{tx}(q,\eta) = - \frac{v^{2}}{\eta} G_{tx}^{tx}(q,\eta)
\en
so
\eq
\sigmaent(q,\omega)
=
\kB^{2}v^{2}\beta^{3}
\int \dif \eta \;
\frac{i}{\omega+i\epsilon-\eta}
\left (1-\me^{\beta\hbar \eta} \right )
\frac{1}{\beta\hbar\eta}
G_{tx}^{tx}(q,\eta)
\:.
\en
In the uniform limit, $q\rightarrow 0$,
\eq
\lim_{q\rightarrow 0}
G_{tx}^{tx}(q,\eta)
= \delta(\eta)
\expvalequil{H_{0}\, T_{x}^{x}(x,t)}
= -\delta(\eta) \partialby\beta \expvalequil{T_{x}^{x}(x,t)}
\en
so
\eq
\sigmaent(\omega)
=\lim_{q\rightarrow 0}
\sigmaent(q,\omega) 
=
\kB^{2}v^{2}\beta^{3}
\frac{i}{\omega+i\epsilon}
\partialby\beta \expvalequil{T_{x}^{x}(x,t)}
\en
The equilibrium entropy density is
(see (\ref{eq:Sdensity})):
\eq
\Sdensity =
\kB \beta^{2} \partialby\beta
\expvalequil{T_{x}^{x}(x,t)}
\en
so
\eq
\sigmaent(\omega)
= 
\frac{i \kB \beta v^{2}\Sdensity}{\omega}
\:.
\en
The thermal conductivity is
\eq
\kappa(\omega,T)  = \Real( T \sigmaent(\omega)) =
\pi v^{2} \Sdensity \delta(\omega)
\:.
\en

\newpage
\bibliographystyle{spmpsci}       
\bibliography{entropy}

\begin{thebibliography}{10}
\providecommand{\url}[1]{{#1}}
\providecommand{\urlprefix}{URL }
\expandafter\ifx\csname urlstyle\endcsname\relax
  \providecommand{\doi}[1]{DOI~\discretionary{}{}{}#1}\else
  \providecommand{\doi}{DOI~\discretionary{}{}{}\begingroup
  \urlstyle{rm}\Url}\fi

\bibitem{AFFLECK1986}
Affleck, I.: Universal term in the free energy at a critical point and the
  conformal anomaly.
\newblock Phys. Rev. Lett. \textbf{56}, 746--748 (1986)

\bibitem{AL1}
Affleck, I., Ludwig, A.W.: Universal noninteger ''ground state degeneracy'' in
  critical quantum systems.
\newblock Phys. Rev. Lett. \textbf{67}, 161 (1991)

\bibitem{Benioff1}
Benioff, P.: Quantum mechanical hamiltonian models of turing machines.
\newblock J. Stat. Phys. \textbf{29}, 515--546 (1982)

\bibitem{Benioff2}
Benioff, P.: Quantum mechanical hamiltonian models of turing machines that
  dissipate no energy.
\newblock Phys. Rev. Lett. \textbf{48}, 1581--1585 (1982)

\bibitem{Bennett}
Bennett, C.: Logical reversibility of computation.
\newblock IBM J. Research and Development \textbf{17}, 525--532 (1973)

\bibitem{Doyon}
Bernard, D., Doyon, B.: Conformal field theory out of equilibrium: a review.
\newblock Journal of Statistical Mechanics: Theory and Experiment
  \textbf{2016}(6), 064,005 (2016)

\bibitem{BLOTECARDYNIGHTINGALE}
Bl\"ote, H., Cardy, J., Nightingale, M.: Conformal invariance, the central
  charge, and universal finite-size amplitudes at criticality.
\newblock Phys. Rev. Lett. \textbf{56}, 742--745 (1986)

\bibitem{Calabreseetal}
Calabrese, P., Essler, F.H.L., Mussardo, G.: Quantum integrability in out of
  equilibrium systems.
\newblock Journal of Statistical Mechanics: Theory and Experiment
  \textbf{2016}(6), 064,001 (2016)

\bibitem{CFL}
Cappelli, A., Friedan, D., Latorre, J.I.: C theorem and spectral
  representation.
\newblock Nucl. Phys. \textbf{B352}, 616--670 (1991)

\bibitem{Cardybook}
Cardy, J.: Scaling and Renormalization in Statistical Physics.
\newblock Cambridge University Press, Cambridge (1996)

\bibitem{QuantumCircuits8}
Chamon, C., Oshikawa, M., Affleck, I.: Junctions of three quantum wires and the
  dissipative hofstadter model.
\newblock Phys. Rev. Lett. \textbf{91}, 206,403 (2003)

\bibitem{DiFran}
DiFrancesco, P., Mathieu, P., Senechal, D.: Conformal Field Theory.
\newblock Springer-Verlag, New York (1997)

\bibitem{dixon:1988}
Dixon, L., Ginsparg, P., Harvey, J.: Beauty and the beast:superconformal
  symmetry in a monster module.
\newblock Comm.~Math.~Phys. \textbf{119}, 221--241 (1988)

\bibitem{MONSTER1988}
Frenkel, I., Lepowsky, J., Meurman, A.: Vertex Operator Algebras and the
  Monster.
\newblock Pure and Applied Mathematics Volume 134, Academic Press, San Diego
  (1988)

\bibitem{MONSTER1984}
Frenkel, I.B., Lepowsky, J., Meurman, A.: A natural representation of the
  fischer-griess monster with the modular function j as a character.
\newblock Proc. Nat. Acad. Sci. USA \textbf{81}, 3256--3260 (1984)

\bibitem{MONSTER1983}
Frenkel, I.B., Lepowsky, J., Meurman, A.: A moonshine module for the monster.
\newblock In: J.~Lepowsky, S.~Mandelstam, I.~Singer (eds.) Vertex Operators in
  Mathematics and Physics - Proceedings of a Conference November 10-17, 1983,
  no.~3 in Publications of the Mathematical Sciences Research Institute, pp.
  231--273. Springer, New York (1985)

\bibitem{EntFlowI}
Friedan, D.: Entropy flow in near-critical quantum circuits (2005).
\newblock {arXiv:cond-mat/0505084v1}

\bibitem{DFEntropyFlowII}
Friedan, D.: Entropy flow through near-critical quantum junctions (2005).
\newblock {arXiv:cond-mat/0505085v1}

\bibitem{DFEntropyFlowIIJStatPhys}
Friedan, D.: Entropy flow through near-critical quantum junctions.
\newblock J. Stat. Phys.  (2017).
\newblock {doi:10.1007/s10955-017-1752-8. arXiv:cond-mat/0505085v2}

\bibitem{FK}
Friedan, D., Konechny, A.: Boundary entropy of one-dimensional quantum systems
  at low temperature.
\newblock Phys. Rev. Lett. \textbf{93}, 030,402 (2004)

\bibitem{FQS1984}
Friedan, D., Qiu, Z., Shenker, S.: Conformal invariance, unitarity, and
  critical exponents in two dimensions.
\newblock Phys. Rev. Lett. \textbf{52}, 1575--1578 (1984)

\bibitem{FQS1983}
Friedan, D., Qiu, Z., Shenker, S.: Conformal invariance, unitarity and two
  dimensional critical exponents.
\newblock In: J.~Lepowsky, S.~Mandelstam, I.~Singer (eds.) Vertex Operators in
  Mathematics and Physics - Proceedings of a Conference November 10-17, 1983,
  no.~3 in Publications of the Mathematical Sciences Research Institute, pp.
  419--449. Springer, New York (1985)

\bibitem{FS1986}
Friedan, D., Shenker, S.: Supersymmetric critical phenomena and the two
  dimensional gaussian model  (1986).
\newblock {preprint, Enrico Fermi Institute, reprinted in {Conformal Invariance
  and Applications to Statistical Mechanics}, eds. C.~Itzykson, H.~Saleur, and
  J.B.~Zuber (World Scientific, Singapore, 1988), pp. 578--579 }

\bibitem{Gibbs}
Gibbs, J.W.: Letter to the secretary of the electrolysis committee of the
  british association for the advancement of science.
\newblock Report Brit. Asso. Adv. Sci. pp. 343--346 (1888).
\newblock Reprinted in {The Collected Works of J. Willard Gibbs,} Yale
  University Press (New Haven, 1928, 1948), vol. 1, pp. 408-412.

\bibitem{KlumperSakai}
Kl\"umper, A., Sakai, K.: The thermal conductivity of the spin-1/2 xxz chain at
  arbitrary temperature.
\newblock Journal of Physics A: Mathematical and General \textbf{35}(9), 2173
  (2002)

\bibitem{Landauer}
Landauer, R.: Irreversibility and heat generation in the computing process.
\newblock IBM J. Research and Development \textbf{3}, 183--191 (1961)

\bibitem{luttinger:1964}
Luttinger, J.M.: Theory of thermal transport coefficients.
\newblock Physical Review \textbf{135}, A1505ÐA1514 (1964)

\bibitem{MacLennan}
McLennan, J.A.: The Formal Statistical Theory of Transport Processes, pp.
  261--317.
\newblock John Wiley \& Sons, Inc. (2007)

\bibitem{orignac:134426}
Orignac, E., Chitra, R., Citro, R.: Thermal transport in one-dimensional spin
  gap systems.
\newblock Phys. Rev. \textbf{B67}(13), 134,426 (2003)

\bibitem{Sachdevbook}
Sachdev, S.: Quantum Phase Transitions.
\newblock Cambridge University Press, Cambridge (1999)

\bibitem{Sachdevarticle}
Sachdev, S.: Quantum phase transitions.
\newblock In: G.~Fraser (ed.) The New Physics For the Twenty-First Century,
  second edn. Cambridge University Press, Cambridge (2005)

\bibitem{Zamolodchikov:1989cf}
Zamolodchikov, A.B.: Thermodynamic bethe ansatz in relativistic models. scaling
  three state potts and lee-yang models.
\newblock Nucl. Phys. \textbf{B342}, 695--720 (1990)

\bibitem{Zubarev}
Zubarev, D.N.: Nonequilibrium Statistical Thermodynamics.
\newblock New York: Consultants Bureau (1974)

\end{thebibliography}

\end{document}